\documentclass[preprint,onecolumn,superscriptaddress,amsmath,amssymb,aps,floatfix,nofootinbib,longbibliography]{revtex4-1}

\usepackage[pdftex]{graphicx}
\usepackage{times}
\usepackage{amsmath}
\usepackage{amssymb}

\begin{document}

\title{From arteries to boreholes: Steady-state response \\ of a poroelastic cylinder to fluid injection}

\author{Lucy C. Auton}
\affiliation{Department of Engineering Science, University of Oxford, Oxford, OX1 3PJ, UK}
\author{Christopher W. MacMinn}
\email{christopher.macminn@eng.ox.ac.uk}
\affiliation{Department of Engineering Science, University of Oxford, Oxford, OX1 3PJ, UK}

\date{\today}


\begin{abstract}
The radially outward flow of fluid into a porous medium occurs in many practical problems, from transport across vascular walls to the pressurisation of boreholes. As the driving pressure becomes non-negligible relative to the stiffness of the solid structure, the poromechanical coupling between the fluid and the solid has an increasingly strong impact on the flow. For very large pressures or very soft materials, as is the case for hydraulic fracturing and arterial flows, this coupling can lead to large deformations and, hence, to strong deviations from a classical, linear-poroelastic response. Here, we study this problem by analysing the steady-state response of a poroelastic cylinder to fluid injection. We consider the qualitative and quantitative impacts of kinematic and constitutive nonlinearity, highlighting the strong impact of deformation-dependent permeability. We show that the wall thickness (thick \textit{vs}. thin) and the outer boundary condition (free \textit{vs}. constrained) play a central role in controlling the mechanics.
\end{abstract}

\maketitle

\section{Introduction}
\label{s:intro}

The radially outward flow of fluid into a porous medium is central to many practical problems in, for example, geomechanics, biophysics, and filtration. In geomechanics, pile driving involves the mechanical expansion of a cylindrical cavity in a fluid-saturated soil, generating large pore pressures in the surrounding medium that gradually relax through consolidation~\citep[\textit{e.g.},][]{randolph1979analytical}. Similarly, fluid injection into boreholes involves the pressurisation of a cylindrical cavity in a soil or rock, driving flow radially outward into the surrounding medium~\citep[\textit{e.g},][]{seth1968transientb, rice1976some, detournay1988poroelastic, sciarra-intjgeomech-2005}. Biophysical applications include injection into subcutaneous tissue~\citep[\textit{e.g},][]{van2012needle} and blood flow through arteries and vascular networks, which have permeable walls~\citep[\textit{e.g},][]{kenyon1979mathematical, jayaraman1983water, klanchar1987modeling, barry1993radial, barry1998effect, skotheim2005physical, reichold2009vascular}. Radially outward flow is also relevant to the design of cylindrical filters~\citep[\textit{e.g},][]{chou2013robust}. In many of these cases, the driving pressure is sufficiently large relative to the stiffness of the solid structure that the poromechanical coupling between the fluid and the solid has an important impact on the flow. Classically, this coupling is described by the iconic theory of linear poroelasticity~\citep[\textit{e.g},][]{biot1941general, wang2000theory}, which combines Darcy's law with linear elasticity in a linearised kinematic framework and is valid for infinitesimal deformations of the solid. However, soft materials such as biological tissues, weak materials such as soils, thin structures such as vasculature, and scenarios involving large injection pressures such as hydraulic fracturing may result in substantial deformations that violate this linear theory. Large deformations are inherently nonlinear from the perspective of kinematics, and typically also result in nonlinear constitutive behaviour such as nonlinear elasticity and deformation-dependent permeability. Recent work in biomechanics and geomechanics, in particular, has focused on capturing the complex material- and application-specific behaviours of tissues and soils~\citep[\textit{e.g.},][]{argoubi-jbiomech-1996, federico-mechmater-2012, tomic-imajapplmath-2014, vuong-compmethapplmecheng-2015, uzuoka-intjnag-2011, song-vzj-2014, borja-compmethapplmecheng-2016}.

Our goal here is to focus on the mechanics of large radial deformations in the context of a simple model problem. We work with relatively generic constitutive laws to avoid obscuring the universal physics of these problems with material-specific behaviour. Historically, uniaxial deformation has been a key model problem for studying the importance of nonlinearity, both mathematically and experimentally \citep[\textit{e.g.,}][]{beavers1981fluidpart2, parker1987steady, hewitt2016flow, macminn2016large}. The uniaxial problem is important for a variety of practical applications; for example, many composite manufacturing processes involve the uniaxial injection of a resin gel or metal melt into a deformable porous matrix~\cite{preziosi1996infiltration}. Mathematically, the uniaxial problem is inherently simple since the flow and deformation fields are strictly one-dimensional and the exact relationship between displacement and porosity is linear~\citep{macminn2016large}. Radial deformations are more challenging despite the fact that the velocity and displacement fields remain one-dimensional, since the stress and strain fields become biaxial and the exact relationship between the porosity and displacement becomes nonlinear.

Radial poroelastic deformations have been studied using linear poroelasticity in the context of both fluid injection or extraction from boreholes~\citep[\textit{e.g},][]{seth1968transientb, rice1976some, detournay1988poroelastic} and arterial blood flow~\citep[\textit{e.g.},][]{kenyon1979mathematical, jayaraman1983water}. Nonlinear effects have attracted interest primarily in the latter case, specifically in the context of fluid flow through artery walls. For example, \citet{klanchar1987modeling} introduced deformation-dependent permeability within a linear poroelastic framework. \citet{barry1993radial} and \citet{barry1998effect} accounted partially for the nonlinear kinematics of large deformations while retaining linear elasticity. In a different context, \citet{macminn2015fluid} developed a rigorous and fully nonlinear model, but for a strictly volumetric constitutive law and assuming constant permeability. None of these previous works explicitly defined or explored the general parameter space for axisymmetric deformations, nor did they systematically assess the relative importance of nonlinear kinematics, nonlinear elasticity, and deformation-dependent permeability.

Here, we consider the axisymmetric deformation of a poroelastic cylinder driven by radially outward fluid flow using a rigorous, fully nonlinear model. We focus, in particular, on the qualitative and quantitative implications of the simplifications of linear poroelasticity, the separate roles of nonlinear kinematics, nonlinear elasticity, and deformation-dependent permeability, and the nontrivial coupling of these with the geometry and boundary conditions. We show that the wall thickness and the outer boundary condition play crucial roles in controlling the mechanics of the problem.

\section{Model Problem}
\label{s:rad}

\begin{figure}[tb]
    \centering
    \includegraphics[width=5in]{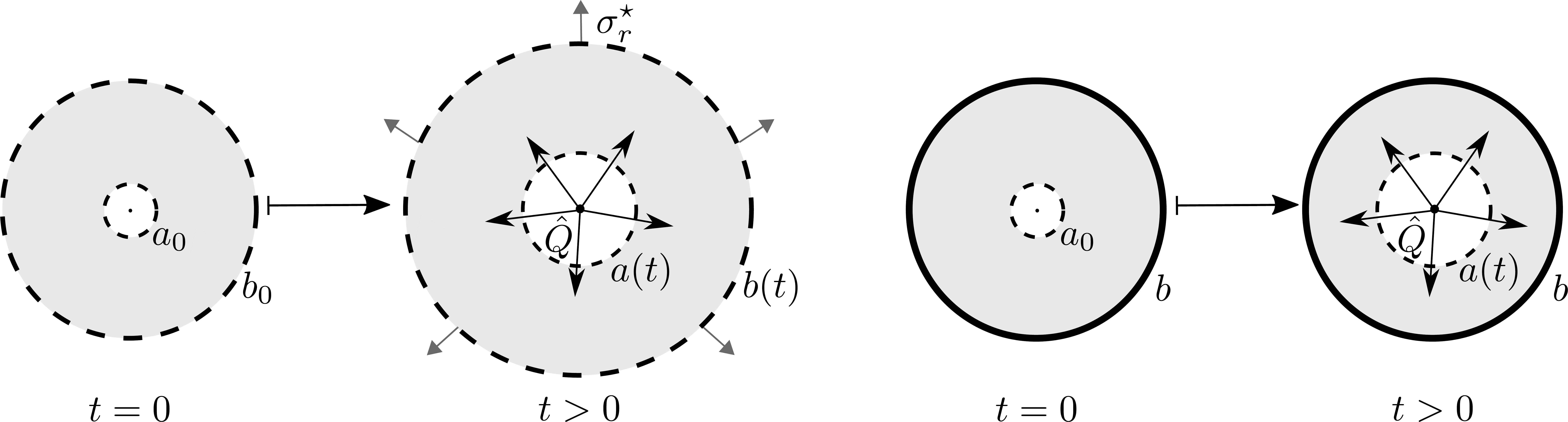}
    \caption{Radially outward fluid flow through a soft porous cylinder of initial inner radius $a_0$ and initial outer radius $b_0$. The inner radius is free to expand, while the outer boundary is either subject to a constant radial effective stress $\sigma_r^\star$ (left) or fixed in place (right). Note that we assume plane strain and adopt the convention of tension being positive. \label{fig:rad} }
\end{figure}

We consider the radially outward injection of fluid from the centre of a porous cylinder of inner radius $a$ and outer radius $b$. We assume axisymmetry and model the 2D annular cross-section, assuming that the material is constrained in the axial direction and is therefore in plane strain. We assume that the inner boundary is mechanically free so that the inner radius $a=a(t)$ expands in response to injection. We assume that the outer boundary is either subject to a constant effective stress $\sigma_r^\star$, in which case the outer radius $b=b(t)$ also expands in response to injection (Fig.~\ref{fig:rad}, left), or that the outer boundary is constrained such that the outer radius $b=b_0$ is fixed (Fig.~\ref{fig:rad}, right). The latter situation is useful for comparison to numerical simulations and experiments~\citep[\textit{e.g.},][]{macminn2015fluid}.

\subsection{Summary of Theory}

Large-deformation poroelasticity is a continuum approach to modelling the interactions of two superposed phases, a porous solid skeleton and an interstitial fluid~\citep[\textit{e.g.},][]{macminn2016large}. We next summarise this theory in the context of axisymmetric flow and deformation.

\subsubsection{Kinematics}

The fluid velocity $\mathbf{v}_f$, solid displacement $\mathbf{u}_s$, and solid velocity $\mathbf{v}_s$ each have only one component, 
\begin{equation}
    \mathbf{u}_s=u_{s}(r,t)\hat{\mathbf{e}}_r, \ \mathbf{v}_s=v_{s}(r,t)\hat{\mathbf{e}}_r, \ \mathrm{and} \ \mathbf{v}_f=v_{f}(r,t)\hat{\mathbf{e}}_r,
\end{equation}
where the subscripts $s$ and $f$ denote quantities related to the solid and to the fluid, respectively, $r$ is the radial coordinate $(a\leq{}r\leq{}b)$, $t$ is time, and $\hat{\mathbf{e}}_r$ is the radial unit vector. We work in an Eulerian (spatial) reference frame, such that the displacement is given by
\begin{equation}
\label{disp}
u_{s} (r,t)= r-R(r,t),
\end{equation}
where $R(r,t)$ denotes the reference position of the material that is located at position $r$ at time $t$. Without loss of generality, we take $u_s(r,0)=0$ such that $R(r,0) = r$---that is, we adopt the initial configuration as the reference configuration. The deformation is fully characterised by the deformation gradient tensor $ \mathbf{F} = (\mathbf{I}-\nabla\mathbf{u_s})^{-1}, $ where $\mathbf{I} $ denotes the identity tensor and $(\cdot)^{-1}$ the inverse. For an axisymmetric deformation, this can be written
\begin{equation}
    \mathbf{F} = \left( \begin{array}{ccc}
    \lambda_r & 0 & 0\\
    0 & \lambda_\theta & 0\\
    0 & 0 & \lambda_z\end{array}\right),
\end{equation}
where $\lambda_r$, $\lambda_\theta$, and $\lambda_z$ are the three principal stretch ratios.\footnote{In general, $\lambda_i^2$ are the eigenvalues of $\mathbf{F}\mathbf{F}^\mathsf{T}$.} For plane strain, these are given by

\begin{equation}\label{stretches}
    \lambda_r = \left(1-\frac{\partial u_{s}}{\partial r}\right)^{-1}, \quad \lambda_\theta = \left(1-\frac{u_{s}}{ r}\right)^{-1}, \quad\text{and}\quad \lambda_z \equiv 1.
\end{equation}
Note that although the displacement field is one dimensional, the state of strain is indeed two dimensional (\textit{i.e.}, both $\lambda_r$ and $\lambda_\theta$ are distinct and nontrivial).

The Jacobian determinant $J\equiv$ det$(\mathbf{F})$ measures the local volume change,
\begin{equation}\label{jac}
    J=\lambda_r\lambda_\theta\lambda_z =\lambda_r\lambda_\theta.
\end{equation}
We assume that the solid and fluid phases are individually incompressible, such that deformation occurs only through rearrangement of the solid skeleton with corresponding changes in the local porosity or fluid fraction, $\phi_f$. This then requires that
\begin{equation}\label{jac2}
    J(r,t)= \frac{1-\phi_{f,0}}{1-\phi_f},
\end{equation}
where $\phi_{f,0}$ is the reference (initial) porosity, which we take to be uniform. Combining Eqs.~(\ref{stretches}--\ref{jac2}), we obtain an explicit nonlinear expression for porosity in terms of displacement,
\begin{equation}\label{eq:phi_to_u}
    \frac{\phi_f-\phi_{f,0}}{1-\phi_{f,0}} =\frac{1}{r}\frac{\partial}{\partial{r}}\left(ru_s-\frac{1}{2}u_s^2\right).
\end{equation}

Conservation of mass for the fluid-solid mixture is given by
 \begin{equation}
 \label{cons}
 \frac{\partial\phi_f}{\partial t} + \frac{1}{r}\frac{\partial}{\partial r} \left(r\phi_fv_{f}\right)=0 \quad \mathrm{and}\quad \frac{\partial\phi_f}{\partial t} - \frac{1}{r}\frac{\partial}{\partial r} \left[r(1-\phi_f) v_{s}\right]=0,
 \end{equation}
where $1-\phi_f$ is the local solid fraction. Conservation of solid volume requires that
\begin{equation}\label{eq:Vol_rad}
    \int_a^b\,2\pi{}r\,(1-\phi_f)\,\mathrm{d}r =\pi\left(b_0^2-a_0^2\right)(1-\phi_{f,0}),
\end{equation}
and it can be shown that Eq.~\eqref{eq:Vol_rad} is identically satisfied by Eq.~(\ref{eq:phi_to_u}), subject to the kinematic boundary conditions $u_s(a,t)= a(t)-a_0$ and $u_s(b,t) = b(t)-b_0$, where $a_0\equiv{}a(0)$ and $b_0\equiv{}b(0)$ denote the initial inner and outer radii, respectively.

\subsubsection{Darcy's Law}

We assume that fluid flows relative to the solid skeleton according to Darcy's law. In the absence of gravity and other body forces, this can be written
\begin{equation}\label{darcy}
    \phi_f(v_{f}-v_{s})=-\frac{k(\phi_f)}{\mu} \frac{\partial p}{\partial r},
\end{equation}
where $\mu$ is the dynamic viscosity of the fluid, $p$ is the fluid (pore) pressure, and $k(\phi_f)$ is the permeability, which we take to be an isotropic function of porosity (see \S\ref{ss:permeability}).

We model injection as a line source at the origin with flow rate per unit length $\hat{Q}(t)$. Thus, Eqs.~(\ref{cons}) can be summed and integrated to give
\begin{equation}\label{Q}
    2\pi r[\phi_fv_{f}+(1-\phi_f)v_{s}] = \hat{Q}(t).
\end{equation}
Combining Eq.~(\ref{Q}) with Eqs.~(\ref{cons}) and (\ref{darcy}), we eliminate $ \phi_s $, $v_{s}$, and $v_{f}$ to obtain 
\begin{subequations}
\begin{equation}\label{governalso}
 \frac{\partial\phi_f}{\partial t} + \frac{1}{r}\frac{\partial}{\partial r} \left(\phi_f\frac{\hat{Q}(t)}{2\pi}-r(1-\phi_f)\frac{k(\phi_f)}{\mu}\frac{\partial p}{\partial r}\right)=0, 
\end{equation}
where along the way we obtain expressions for $v_{f}$ and $v_{s}$, 
\begin{equation} \label{eq:vs}
    v_{f} =\frac{\hat{Q}(t)}{2\pi{}r}-\frac{1-\phi_f}{\phi_f}\frac{k(\phi_f)}{\mu} \frac{\partial{p}}{\partial{r}} \quad \mathrm{and} \quad v_{s}=\frac{\hat{Q}(t)}{2\pi r} +\frac{k(\phi_f)}{\mu}\frac{\partial{p}}{\partial{r}}.
\end{equation}
\end{subequations}
We next link the fluid pressure to the stress in the solid.

\subsubsection{Mechanical equilibrium}

Mechanical equilibrium requires that 
\begin{equation}\label{mecheqb}
    \boldsymbol{\nabla}\cdot\boldsymbol{\sigma} = 0,
\end{equation}
where $\boldsymbol{\sigma}$ is the total stress supported by the fluid-solid mixture, and we neglect inertia as well as the effect of gravity and other body forces. The total stress can be decomposed as 
\begin{equation}
\label{stressful2}
\boldsymbol{\sigma}=\boldsymbol{\sigma'}-p\mathbf{I},
\end{equation}
where Terzaghi's effective stress $\mathbf{\sigma}'$ is the portion of the stress supported through deformation of the solid skeleton, and where we adopt the convention of tension being positive. Equation~(\ref{stressful2}) provides mechanical coupling between the fluid and the solid. Combining Eqs.~(\ref{mecheqb}) and (\ref{stressful2}) leads, for an axisymmetric deformation, to 

\begin{equation}\label{govern}
    \frac{\partial\sigma'_r}{\partial r}+\frac{\sigma'_r-\sigma'_{\theta}}{r}=\frac{\partial p}{\partial r},
\end{equation}
where $\sigma'_r$ and $\sigma'_{\theta}$ are the radial and azimuthal (``hoop'') components of the effective stress, respectively. 

\subsubsection{Linearisation}
\label{lin}

We have now considered kinematics, Darcy's Law, Terzaghi's effective stress, and mechanical equilibrium. The model thus far is exact, assuming only that the fluid and solid constituents are individually incompressible.

The common assumption of infinitesimal deformations leads to classical linear poroelasticity~\citep{wang2000theory, macminn2016large}. This corresponds here to the assumptions that $u_s/r\ll1$ and $ \partial u_s/\partial r \ll1 $. Note that this will clearly be a bad assumption near the inner radius if $u_s$ becomes comparable to $a_0$. Linearising Eqs.~(\ref{eq:phi_to_u}) and (\ref{governalso}) leads to
\begin{equation}\label{eq:phi_to_ulin}
    \quad\frac{\phi_f-\phi_{f,0}}{1-\phi_{f,0}} \approx\frac{1}{r}\frac{\partial}{\partial{r}}\left(ru_s\right) \quad\mathrm{and}\quad \frac{\partial\phi_f}{\partial t} - \frac{1}{r}\frac{\partial}{\partial r} \left(r(1-\phi_{f,0})\frac{k(\phi_{f,0})}{\mu}\frac{\partial p}{\partial r}\right)\approx0,
\end{equation}
respectively. Note that Eq.~(\ref{eq:Vol_rad}) is not identically satisfied by the kinematic expression in Eq.~(\ref{eq:phi_to_ulin}), implying that the linearised model is not rigorously mass conservative. We next consider the constitutive behaviour of the solid.

\subsection{Constitutive laws}

The relationships between stress and strain and between strain and displacement are constitutive laws for the solid skeleton. We assume that the solid deforms elastically, meaning that these relationships are quasi-static (\textit{i.e.}, rate independent) and reversible (\textit{i.e.}, history independent). We investigate the impact of this relationship on the results by considering both linear and nonlinear elasticity laws.

\subsubsection{Hencky Elasticity}

Hencky elasticity is a simple, nonlinear, hyperelastic model that is based on a logarithmic strain measure and provides good agreement with experiments for moderate deformations~\citep{hencky1931law, anand1979h}. In uniaxial compression, Hencky elasticity provides a stiffer response than linear elasticity, with the stress diverging as the thickness of the material approaches zero; in uniaxial tension, Hencky elasticity provides a softer response than linear elasticity, with the stress reaching a maximum and then decaying asymptotically to zero (see Appendix~\ref{supp:Hencky}).

Hencky elasticity has several advantageous properties \cite{bazant1998easy}, including that it reduces to linear elasticity in the limit of infinitesimal deformations and that it uses the same elastic parameters as linear elasticity \cite{xiao2002hencky}. We work here in terms of Lam\'{e}'s first parameter $ \Lambda$ and the $p$-wave or oedometric modulus~$\mathcal{M}$.

For the displacement field given in Eq.~\eqref{disp}, the Hencky strain tensor is 
\begin{equation}\label{eq:H_rad}\renewcommand{\arraystretch}{1.5}
    \boldsymbol{\varepsilon}=\left[
    \begin{array}{ccc}
        \ln{\lambda_r} & 0 & 0 \\
        0 & \ln{\lambda_\theta} & 0 \\
        0 & 0 & 0
    \end{array}\right],
\end{equation}
which again has two nontrivial components since axisymmetric displacement leads to both radial and azimuthal strains. The associated Cauchy effective stress for Hencky elasticity is
\begin{equation} \label{henckconst}
    \boldsymbol{\sigma}^{\prime}=\left[
    \begin{array}{ccc}
        \mathcal{M}\displaystyle\frac{\ln{\lambda_r}}{J} +\Lambda\displaystyle\frac{\ln{\lambda_\theta}}{J} & 0 & 0 \\
        0 & \Lambda\displaystyle\frac{\ln{\lambda_r}}{J} +\mathcal{M}\displaystyle\frac{\ln{\lambda_\theta}}{J} & 0 \\
        0 & 0 & \Lambda\left(\displaystyle\frac{\ln{\lambda_r}+\ln{\lambda_\theta}}{J}\right)
    \end{array}\right].
\end{equation}
On substitution of Eq.~(\ref{henckconst}) into Eq.~(\ref{govern}), we arrive at 
\begin{equation}\label{eq:gradp_to_u_rad_large}\renewcommand{\arraystretch}{1.5}
    \frac{\partial p}{\partial r} =\frac{\partial}{\partial{r}} \left(\mathcal{M}\,\frac{\ln{\lambda_r}}{J}+\Lambda\frac{\ln{\lambda_\theta}}{J}\right) +\frac{\mathcal{M}-\Lambda}{r}\left(\frac{\ln{\lambda_r}}{J}-\frac{\ln{\lambda_\theta}}{J}\right).
\end{equation}
The right-hand side of Eq.~\eqref{eq:gradp_to_u_rad_large} is a function of $u_s$ only. In combination with Eqs.~(\ref{eq:phi_to_u}) and (\ref{governalso}), this then provides a nonlinear partial differential equation (PDE) for $u_s$.

\subsubsection{Linear elasticity}
\label{eq:LinearElasticity}

Linear elasticity combines a linear relationship between strain and displacement with a linear relationship between stress and strain. The linear (small or infinitesimal) strain tensor is
\begin{equation}\label{eq:eps_small_rad}\renewcommand{\arraystretch}{1.5}
    \boldsymbol{\varepsilon}=\left[
    \begin{array}{ccc}
        \displaystyle\frac{\partial{u_s}}{\partial{r}} & 0 & 0 \\
        0 & \displaystyle\frac{u_s}{r} & 0 \\
        0 & 0 & 0
    \end{array}\right]
\end{equation}
with the associated linear stress tensor
\begin{equation}\label{eq:sigma_small_rad}\renewcommand{\arraystretch}{1.5}
    \boldsymbol{\sigma}^{\prime}=\left[
    \begin{array}{ccc}
        \mathcal{M} \displaystyle\frac{\partial{u_s}}{\partial{r}} + \Lambda \displaystyle\frac{u_s}{r} & 0 & 0 \\
        0 & \Lambda \displaystyle\frac{\partial{u_s}}{\partial{r}} +\mathcal{M} \displaystyle\frac{u_s}{r} & 0 \\
        0 & 0 & \Lambda\left( \displaystyle\frac{\partial{u_s}}{\partial{r}}+\displaystyle\frac{u_s}{r}\right)
    \end{array}\right].
\end{equation}
On substitution of Eq.~(\ref{eq:sigma_small_rad}) into Eq.~(\ref{govern}), we obtain 

\begin{equation}\label{eq:p_to_u_small_rad}
    \frac{\partial{p}}{\partial{r}} =\frac{\partial}{\partial{r}} \left[\mathcal{M}\,\frac{\partial{u_s}}{\partial{r}}+\Lambda\frac{u_s}{r}\right] +\frac{\mathcal{M}-\Lambda}{r}\left(\frac{\partial{u_s}}{\partial{r}}-\frac{u_s}{r}\right)=\mathcal{M}\,\frac{\partial}{\partial{r}} \left[\frac{1}{r}\frac{\partial}{\partial{r}}\left(ru_s\right)\right].
\end{equation}
Linear elasticity is in some sense an idealised constitutive behaviour that most materials will approximately follow for infinitesimal deformations, and from which most materials will deviate as deformations become finite. For example, Hencky elasticity reduces to linear elasticity for infinitesimal deformations; that is, Eqs.~(\ref{eq:H_rad}) and (\ref{henckconst}) reduce to Eqs.~(\ref{eq:eps_small_rad}) and Eq.~(\ref{eq:sigma_small_rad}), respectively, for $u_s/r \ll 1$ and $\partial u_s/\partial r \ll1 $. Alternatively, linear elasticity can instead be viewed as an exact constitutive law for an idealised material, for which it would be valid for arbitrarily large deformations.

Equation~(\ref{eq:p_to_u_small_rad}) can be combined with Eqs.~(\ref{eq:phi_to_u}) and (\ref{governalso}) to provide a PDE for $u_s$. In what follows, we use ``Hencky elasticity'' to refer to Eqs.~(\ref{eq:H_rad}--\ref{eq:gradp_to_u_rad_large}) and ``linear elasticity'' to refer to Eqs.~(\ref{eq:eps_small_rad}--\ref{eq:p_to_u_small_rad}).

\subsubsection{Linear poroelasticity}

We now combine linearised kinematics (\S\ref{lin}) with linear elasticity (\S\ref{eq:LinearElasticity}). This then allows us to write Eq.~(\ref{eq:p_to_u_small_rad}) directly in terms of $\phi_f$ using Eq.~(\ref{eq:phi_to_ulin}),
\begin{equation}\label{eq:p_to_phi_small_rad}
    \frac{\partial p}{\partial r} \approx\mathcal{M}\,\frac{\partial}{\partial{r}}\left(\frac{\phi_f-\phi_{f,0}}{1-\phi_{f,0}}\right).
\end{equation}
Equation~(\ref{eq:phi_to_ulin}) can then be rewritten as a linear second-order parabolic PDE for $\phi_f$.

\subsection{Permeability Laws}
\label{ss:permeability}

The solid skeleton deforms through rearrangement of the pore structure, leading to changes in the porosity. This is then likely to alter the permeability of the material. For infinitesimal deformations, this effect is second-order in the deformation, and is therefore typically neglected. We consider the impact of this simplification by comparing results for constant permeability with results for deformation-dependent permeability. As in \citet{macminn2016large}, we adopt a normalised Kozeny-Carman formula,
\begin{equation}
    k(\phi_f) =k_0\frac{(1-\phi_{f,0})^2}{\phi_{f,0}^3}\frac{\phi_f^3}{(1-\phi_f)^2},
\end{equation}
where $k_0\equiv k(\phi_{f,0})$ is the reference permeability. Although not quantitatively appropriate for all materials, this relation captures the important qualitative behaviour that $k(\phi_f)$ vanishes as $\phi_f$ vanishes and $k(\phi_f)$ diverges as $\phi_f$ tends to one.

Note that many materials have a naturally anisotropic permeability. In addition, anisotropic deformations may lead to the emergence of anisotropic permeability. For example, fluid flow through the walls of a porous cylinder leads to compression in the radial direction and stretching in the azimuthal direction, which might be expected to reduce the azimuthal permeability while enhancing the radial permeability. We neglect natural anisotropy here for simplicity, and induced anisotropy is irrelevant under the requirement of axisymmetry.

\subsection{Initial State and Boundary Conditions}
\label{ss:BCs}

Before injection, the porosity is uniform, $\phi_f(r,0)=\phi_{f,0}$, the fluid and the solid are at rest, $v_f(r,0)=v_s(r,0)=0$, and the material is relaxed, $\sigma^\prime_r(r,0)=\sigma^\prime_\theta(r,0)=0$. We take this initial state to be the reference state, such that $u_s(r,0)=0$.

\subsubsection{Injection}

For $t>0$, we assume that fluid is injected from the origin either at an imposed constant volume flow rate per unit length $\hat{Q}$ or via an imposed constant pressure drop $\Delta{p}\equiv{}p(a,t)-p(b,t)$. It is straightforward to enforce the former condition since $\hat{Q}$ appears explicitly in the PDE. Enforcing the latter condition is less straightforward (see \S{}B of the Appendix).

\subsubsection{Inner boundary}

The inner boundary is mechanically free, thus the normal effective stress must vanish. The inner boundary is also a material boundary. Hence, the appropriate mechanical and kinematic conditions are
\begin{equation} \label{innerBC}
    \sigma'_r(a,t)=0, \quad u_{s}(a,t)=a(t)-a_0, \quad \mathrm{and} \quad v_s(a,t)=\frac{\partial{u_s}}{\partial{t}}\Big|_{r=a} = \frac{\mathrm{d}a}{\mathrm{d}t}.
\end{equation}

\subsubsection{Outer boundary}

We consider two distinct sets of conditions at the outer boundary. In both cases, we assume without loss of generality that the fluid pressure vanishes at the outer boundary,
\begin{equation} \label{p}
    p(b,t)=0.
\end{equation}

If the outer boundary is subject to an applied effective stress, then this is a moving boundary. The appropriate mechanical and kinematic conditions are
\begin{equation}\label{outerBCa}
    \sigma_r^\prime(b,t) = \sigma_r^\star, \quad u_{s}(b,t)=b(t)-b_0, \quad \mathrm{and} \quad v_s(b,t)=\frac{\partial{u_s}}{\partial{t}}\Big|_{r=b} = \frac{\mathrm{d}b}{\mathrm{d}t}.
\end{equation}
Three conditions are required because the outer radius $b(t)$ is unknown, and must be determined as part of the solution.

Alternatively, if the outer boundary is constrained such that its position is fixed, then the appropriate conditions are
\begin{equation} \label{outerBCb}
    u_s(b,t)=0 \quad \mathrm{and} \quad v_s(b,t)=\frac{\partial{u_s}}{\partial{t}}\Big|_{r=b}=0.
\end{equation}
This scenario requires only two conditions because the outer radius $b$ is fixed and known. The normal component of the effective stress at the outer boundary $\sigma_r^\prime(b,t)$ is unknown, but does not need to be determined as part of the solution.

Conditions~(\ref{outerBCb}) are convenient for comparison with experiments and numerical simulations (\emph{e.g.}, \cite{macminn2015fluid}), and are relevant to industrial applications such as filtration. Conditions~(\ref{outerBCa}) are likely to be more relevant to biomedical and geotechnical applications.

\subsubsection{Linearised boundary conditions}

For the kinematically rigorous models, conditions at the inner and outer boundaries (Eqs.~\ref{innerBC}--\ref{outerBCb}) are applied at $a(t)$ and $b(t)$, respectively. For the kinematically linearised models, these are instead applied at $a_0$ and $b_0$, respectively (\textit{e.g.}, $\sigma^\prime_r(a,t)=0 \,\mapsto\, \sigma^\prime_r(a_0,t)\approx{}0$).

\subsection{Non-dimensionalisation and parameters}
\label{nondim}

To proceed, we non-dimensionalise via the scaling
\begin{equation}
    \tilde{r} = \frac{r}{b_0}, \ \tilde{u}_s=\frac{u_s}{b_0}, \ \tilde{a}=\frac{a}{b_0},\ \tilde{b}=\frac{b}{b_0}, \ \tilde{\sigma}_i^\prime=\frac{{\sigma_i}^\prime}{\mathcal{M}}, \ \tilde{t}=\frac{t}{T_\mathrm{pe}}, \ \tilde{p} = \frac{p}{\mathcal{M}},
\end{equation}
where $T_\mathrm{pe}\equiv{} b_0^2\mu/k_0\mathcal{M}$ is the characteristic poroelastic timescale. We can then rewrite Eq.~(\ref{governalso}) in dimensionless form,
\begin{equation} \label{complicated}
    \frac{\partial\phi_f}{\partial \tilde{t}} + \frac{1}{\tilde{r}}\frac{\partial}{\partial \tilde{r}} \left(\phi_fq(\tilde{t})-\tilde{r}(1-\phi_f)\tilde{k}(\phi_f)\frac{\partial \tilde{p}} {\partial \tilde{r}}\right)=0,
\end{equation}
where $\tilde{k}(\phi_f) = k(\phi_f)/k_0$. Injection is characterised either by a fixed dimensionless flow rate $q$ or by a fixed dimensionless pressure drop $\Delta{\tilde{p}}$,
\begin{equation}
    q\equiv{}\frac{\mu \hat{Q}}{2\pi k_0 \mathcal{M}} \quad\mathrm{or}\quad \Delta{\tilde{p}}\equiv{}\frac{\Delta{p}}{\mathcal{M}},
\end{equation}
where, in the latter case, $q(\tilde{t})$ must be calculated from $\Delta{\tilde{p}}$ as part of the solution. Both of these quantities compare the characteristic pressure due to injection with the characteristic elastic stiffness of the material. The model is additionally characterised by the value of $\phi_{f,0}$ and three other dimensionless parameters:
\begin{equation}
    \Gamma \equiv{}\frac{\Lambda}{\mathcal{M}}, \quad \tilde{a}_0 \equiv{}\frac{a_0}{b_0}, \quad\mathrm{and}\quad \tilde{\sigma}_r^\star\equiv{}\frac{\sigma_r^\star}{\mathcal{M}},
\end{equation}
where $\Gamma$ compares the bulk modulus to the shear modulus ($\Gamma \in [-1/2,1]$, where $\Gamma=1$ corresponds to an incompressible material).

We work in dimensionless quantities from here onwards; hence, we drop the tildes for convenience.

\subsection{Summary of models}
\label{summary}

Thus far, we have developed several different models for the response of a poroelastic cylinder to radially outward flow by considering two different representations of the kinematics (linearised and rigorous), two different elasticity laws (linear and Hencky), and two different permeability laws (constant and Kozeny-Carman). We categorise these models as linear ``L'' (linearised kinematics with linear elasticity), quasi-linear ``Q'' (rigorous kinematics with linear elasticity), and nonlinear ``N'' (rigorous kinematics with Hencky elasticity). For each of these, we consider both constant ``$k_0$'' and Kozeny-Carman ``$k_\mathrm{KC}$'' permeability. We then have six combinations: L-$k_0$, L-$k_\mathrm{KC}$, Q-$k_0$, Q-$k_\mathrm{KC}$, N-$k_0$, and N-$k_\mathrm{KC}$. Note that L-$k_0$ is classical linear poroelasticity and N-$k_\mathrm{KC}$ is fully nonlinear poroelasticity; the other four models are intermediate between these two extremes. Note also that we do not combine linearised kinematics with Hencky elasticity because this is asymptotically inconsistent; linearising the kinematics requires that $u_s/r \ll$ and $\partial u_s/\partial r \ll1$, under which assumptions Hencky elasticity reduces to linear elasticity.

\section{Steady-state solutions}\label{ststsolns}

We now seek solutions to the above models at steady state, for which the fluid velocity is steady ($\partial v_f/\partial t=0$) and the solid is stationary ($v_s=0$). Combining Eqs.~(\ref{cons}), (\ref{darcy}), and (\ref{govern}), we have
\begin{equation} \label{cauchy}
    \frac{\mathrm{d}\sigma_r^\prime}{\mathrm{d}r} +\frac{\sigma_r^\prime-\sigma_\theta^\prime}{r} =\frac{\mathrm{d}p}{\mathrm{d}r}=-\frac{q}{rk(\phi_f)},
\end{equation}
where $\phi_f=\phi_f[u_s(r)]$. Combining this with an elasticity law, a permeability law, and a kinematic relationship between $u_s$ and $\phi_f$ then leads to a second-order ODE in $u_s$ for all models. For linear elasticity (L~and~Q~models), we combine Eq.~(\ref{cauchy}) with Eq.~(\ref{eq:sigma_small_rad}) to arrive at
\begin{equation}\label{a}
    \frac{\mathrm{d}^2u_s}{\mathrm{d} r^2} + \frac{1}{r}\frac{\mathrm{d}u_s}{\mathrm{d}r}-\frac{u_s}{r^2}=-\frac{q}{rk[\phi_f(u_s)]}.
\end{equation}
For Hencky elasticity (N models), we combine Eq.~(\ref{cauchy}) with Eq.~(\ref{henckconst}) to arrive at 
\begin{equation}\label{odeelastreg}
    \frac{\mathrm{d}^2u_{s}}{\mathrm{d}r^2} =\frac{\left(1-\lambda_\theta/\lambda_r\right)[\ln (\lambda_r) +\Gamma\ln(\lambda_\theta)-\Gamma]+(1-\Gamma)\ln\left(\lambda_\theta/\lambda_r\right)-q\lambda_r\lambda_\theta/k[\phi_f(u_s)]}{\lambda_r r \{1-[\ln(\lambda_r)+\Gamma\ln(\lambda_\theta)]\}},
\end{equation}
where the stretches are defined in Eq.~(\ref{stretches}). Note that Eqs.~(\ref{a}) and (\ref{odeelastreg}) are valid for any permeability law, boundary conditions, and treatment of kinematics.

Thus, we have a boundary value problem (BVP) comprising a second-order ODE (Eq.~\ref{a} or Eq.~\ref{odeelastreg}) with two constraints at the inner boundary (Eqs.~\ref{innerBC}) and either three or four constraints at the outer boundary, depending on whether the outer boundary is fixed (Eqs.~\ref{p} and \ref{outerBCb}) or not (Eqs.~\ref{p} and \ref{outerBCa}). For the L-$k_0$ and Q-$k_0$ models, the ODE (Eq.~\ref{a}) can be solved analytically (see Appendix~C). For the L-$k_0$ model, this provides the full solution to the problem. For the Q-$k_0$ model, it remains to solve an implicit algebraic system for $a$ and, depending on the outer boundary condition, for $b$. This can be implemented with standard numerical root-finding techniques. For the other four models, the ODE cannot be solved analytically and we instead solve it numerically using a Chebyshev spectral collocation method, as described in \S\ref{pseudo}.

\subsection{Injection}

An imposed flow rate $q$ will lead to a steady-state pressure drop $\Delta{p}$. The latter is not needed as part of the solution, but can be calculated readily via the integration of Eq.~(\ref{eq:vs}), giving
\begin{equation}\label{eq:q_to_p}
    \Delta{p}=q\,\int_a^b\,\frac{1}{rk(\phi_f)}\,\mathrm{d}r.
\end{equation}
In contrast, an imposed pressure drop $\Delta{p}$ will lead to a steady-state flow rate $q$ that must be calculated as part of the solution by rearranging Eq.~(\ref{eq:q_to_p}). For constant permeability, this relationship becomes
\begin{equation}\label{q}
    \Delta{p}=q\ln(b/a).
\end{equation}
Everything else being fixed, the same steady state can therefore be achieved by imposing either $q$ or $\Delta{p}$. Clearly, the geometry and boundary conditions will have a strong impact on the relationship between $q$ and $\Delta{p}$. We explore this relationship in the next section.

\subsection{Numerical solution via Chebyshev spectral collocation}
\label{pseudo}

When the ODE cannot be solved analytically, it must instead be integrated numerically as a BVP. Here, we use a direct method based on Chebyshev spectral collocation (\textit{i.e.}, a Chebyshev pseudospectral method) \citep[\textit{e.g.},][]{trefethen2000spectral, piche2007solving, bjornaraa2013pseudospectral}. That is, we solve the BVP and all constraints simultaneously using a dense Chebyshev-pseudospectral differentiation matrix and Newton iteration (see Appendix~\ref{supp:Chebyshev}). This approach is robust and accurate, and also allows for the straightforward incorporation of additional unknowns and constraints, such as solving the problem for an imposed pressure drop $\Delta{p}$ rather than for an imposed flow rate $q$. We generate the differentiation matrices using the suite of \verb+MATLAB+ functions provided by \citet{weideman2000matlab}.

\section{Results}

We have developed steady-state solutions for six different models, each for two distinct outer boundary conditions --- a fixed outer boundary (``constrained'') and an applied effective stress $\sigma_r^\star$ at the outer boundary (see \S\ref{summary}). As described in \S\ref{nondim}, these models are characterised by five dimensionless parameters: $\Gamma$, a ratio of elastic constants; $\phi_{f,0}$, the initial porosity; $a_0$, the ratio of the initial inner radius to the initial outer radius; $\sigma_r^\star$, the applied effective stress; and either $q$, the flow rate, or $\Delta{p}$, the pressure drop. To focus on the impact of model choice, boundary conditions, and geometry, we adopt fixed values of $\Gamma=0.4$ and $\phi_{f,0}=0.5$ throughout the rest of the paper. Varying these two parameters across a moderate range of typical values does not lead to dramatic qualitative differences in the resulting behaviour. Similarly, we fix $\sigma_r^\star =0$ (``unconstrained'') for simplicity.

\subsection{Model comparison}

In this section, we compare and contrast the six models for the two different boundary conditions (unconstrained and constrained) in the context of two end-member geometries: a thick-walled cylinder (Fig.~\ref{fig:Thick}) and a thin-walled cylinder (Fig.~\ref{fig:Thin}). This gives us a preliminary sense for how the geometry impacts the mechanics, which is in turn the focus of \S\ref{QL}.

\subsubsection{Unconstrained thick-walled cylinder}

\begin{figure}[tb]
    \centering\includegraphics[width=5in]{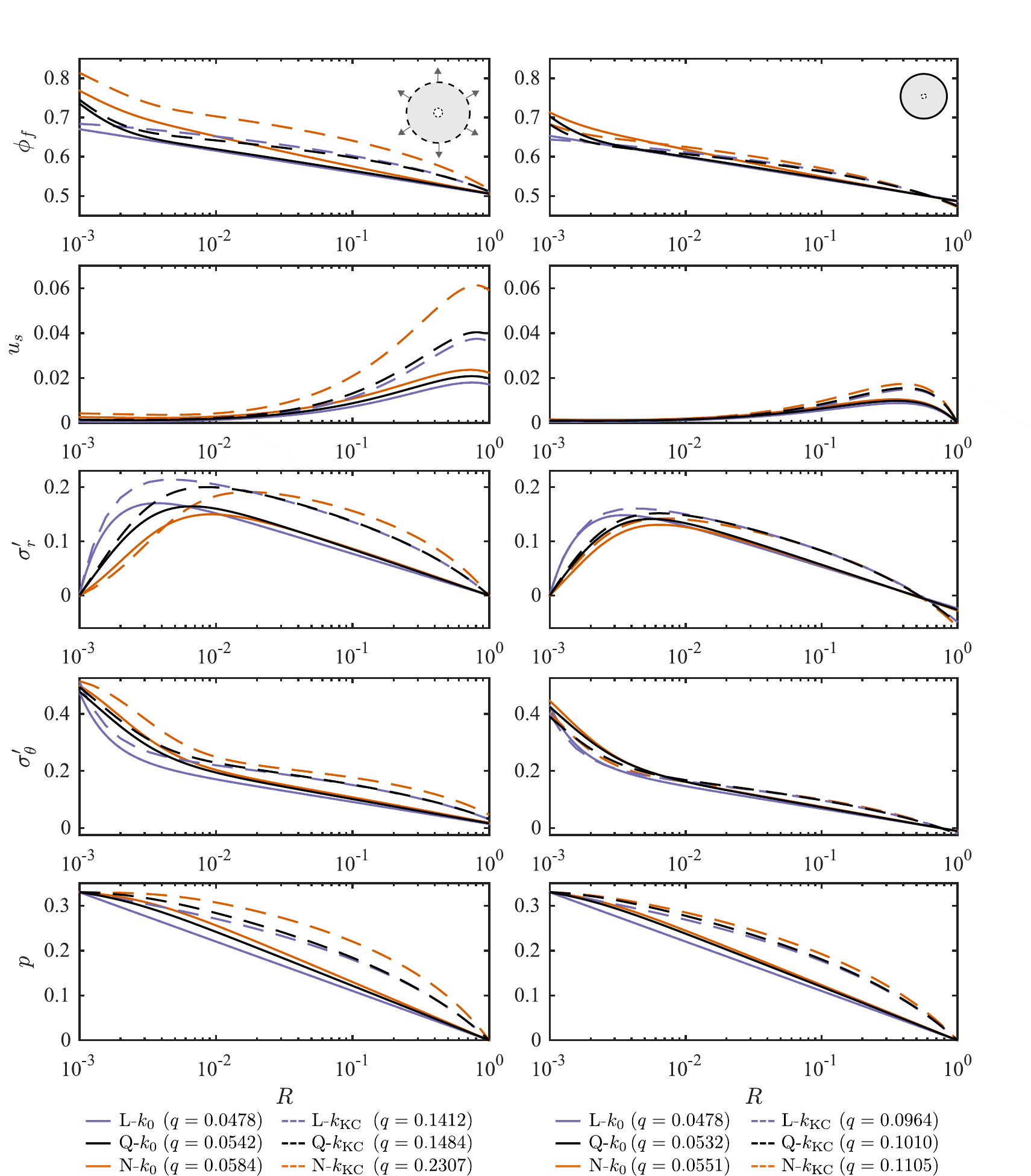}
    \caption{Six models at steady state for a thick-walled cylinder ($a_0=10^{-3}$). We consider an unconstrained cylinder (left column) and a constrained cylinder (right column), both for flow driven by an imposed pressure drop $\Delta{p}=0.33$. For clarity, we plot the results against the Lagrangian coordinate $R(r,t)=r-u_{s}$ and on a logarithmic horizontal scale. The unconstrained and constrained cylinders exhibit very similar behaviour, implying that the distinction between these two outer boundary conditions is unimportant when the walls are very thick (\textit{i.e.}, for small $a_0$). Additionally, note that in this case the permeability law has a stronger impact than the elasticity law or the treatment of the kinematics. \label{fig:Thick} }
\end{figure}

In Fig.~\ref{fig:Thick}, we consider a thick-walled cylinder for flow driven by an imposed pressure drop of $\Delta{p}=0.33$. For an unconstrained thick-walled cylinder (left column), the predictions of all models are qualitatively similar. The porosity $\phi_f$ (top row), azimuthal effective stress $\sigma_\theta^\prime$ (fourth row), and pressure $p$ (last row) all have maxima at the inner boundary and decrease monotonically from left to right. The porosity remains everywhere greater than $\phi_{f,0}$, the azimuthal effective stress is strictly tensile, and the pressure drops from $p(a,t)=\Delta{p}=0.33$ to $p(b,t)=0$ by construction. Additionally, the pressure profile is strongly nonlinear for the $k_\mathrm{KC}$ models, but closer to classical linear poroelasticity (L-$k_0$) for the $k_0$ models. In contrast to the behaviour of these quantities, the displacement $u_s$ (second row) and the radial effective stress $\sigma_r^\prime$ (third row) are non-monotonic. The displacement has an interior maximum that is located in roughly the same place for all models. The radial effective stress vanishes at the inner and outer boundaries by construction. Between these limits, it is purely tensile with an interior maximum, with the location of this maximum depending strongly on model choice.

\subsubsection{Constrained thick-walled cylinder}

For the same pressure drop, a constrained thick-walled cylinder (Fig.~\ref{fig:Thick}, right column) exhibits a strikingly similar behaviour to that of the unconstrained cylinder. The maximum in porosity at the inner boundary is lower than for the unconstrained cylinder, and the porosity now drops slightly below $\phi_{f,0}$ at the outer boundary where the material is slightly compressed. The displacement is qualitatively similar, but a factor of 2--3 smaller than in the unconstrained case. The radial and azimuthal effective stresses are now both slightly compressive at the outer boundary. This comparison between the unconstrained and constrained cylinders supports the intuition that the difference between these two cases becomes unimportant for thick walls (\textit{i.e.,} $a_0\ll{}1$).

In all of the cases shown in Fig.~\ref{fig:Thick}, the flow is driven by the same imposed pressure drop of $\Delta{p}=0.33$. In addition to the above differences between the six models and the two boundary conditions, each of these twelve cases will result in a different flow rate\footnote{Except for the L-$k_0$ model, for which $q$ is independent of the boundary condition.} $q$ (see legend, bottom of Fig.~\ref{fig:Thick}). In all cases, $q$ is lower for the constrained cylinder than for the unconstrained cylinder (again, except for the L-$k_0$ model). This is because the inner radius of the constrained cylinder always expands less than that of the unconstrained cylinder, and $q$ is very sensitive to the inner radius (Eq.~\ref{eq:q_to_p}); the constrained cylinder is also slightly compressed against the outer boundary, which reduces its permeability in the $k_\mathrm{KC}$ models, amplifying the reduction in $q$.

All of the $k_0$ models produce quantitatively similar values of $q$. For each, $q$ differs by only a few percent between the two boundary conditions; between the $k_0$ models for the same boundary condition, $q$ differs by about 10--20\%. By far the largest difference is between the corresponding $k_0$ and $k_\mathrm{KC}$ models, where the $k_\mathrm{KC}$ model produces a value of $q$ that is roughly 2--4 times larger than the corresponding $k_0$ model. The permeability law makes a great difference since large deformations of a thick-walled cylinder lead to large and nonuniform changes in porosity. This substantial change in porosity leads to a substantial change in permeability for the $k_\mathrm{KC}$ models, but has no impact on the $k_0$ models. This effect leads to higher values of $q$ for the $k_\mathrm{KC}$ models because the average porosity is in all cases larger than $\phi_{f,0}$, so the permeability increases. Comparing the N models to the Q models, and the Q models to the L models, reveals that both rigorous kinematics and nonlinear elasticity also lead to higher values of $q$ relative to their linearised counterparts. However, these effects are noticeably weaker than the impact of changing the permeability law. Given that the values of $q$ vary so widely, it is surprising that the behaviour illustrated in Fig.~\ref{fig:Thick} is otherwise so similar across the models and boundary conditions.

\subsubsection{Unconstrained thin-walled cylinder}

\begin{figure}[tb]
    \centering
    \includegraphics[width=5in]{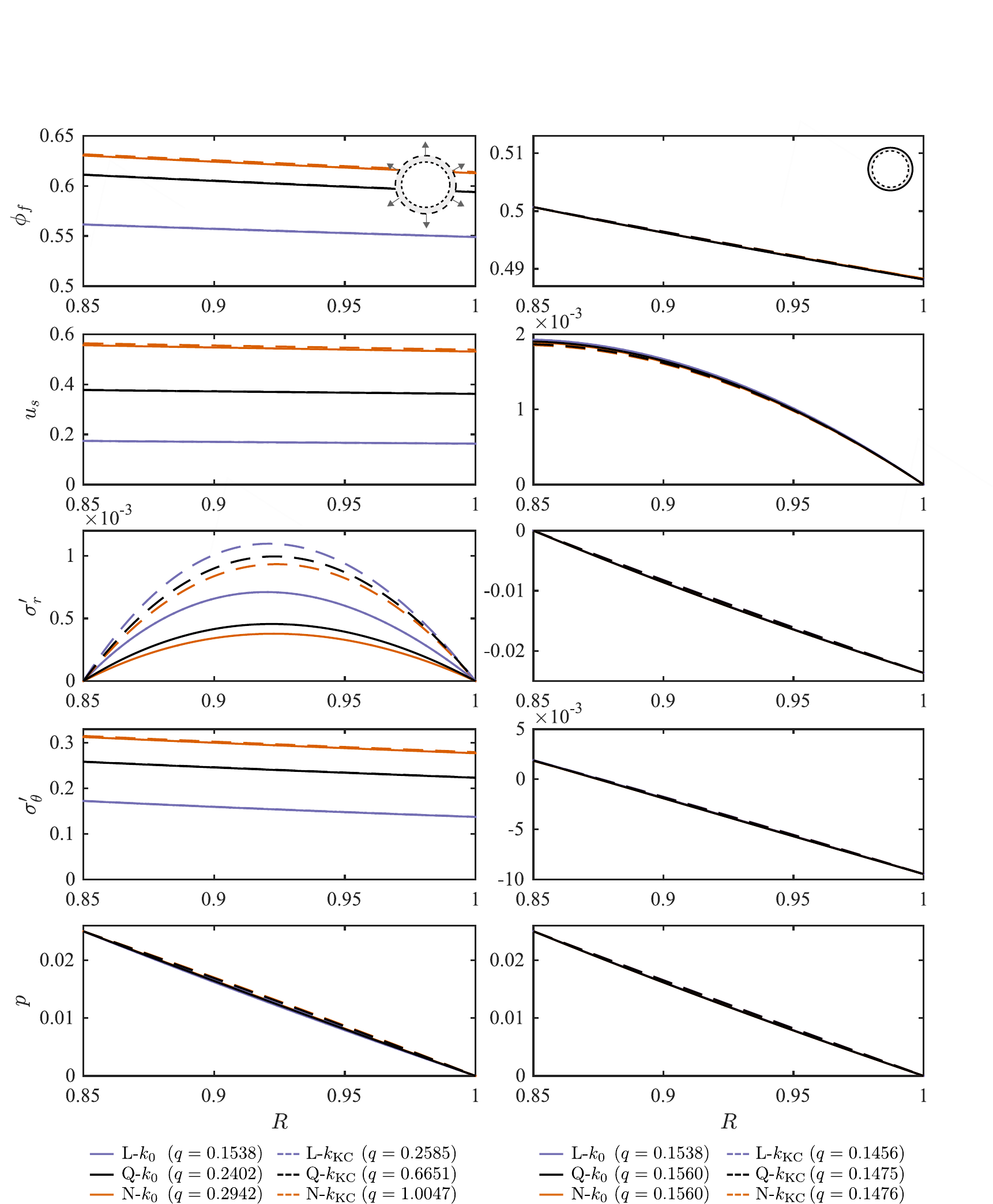}
    \caption{Six models at steady state for a thin-walled cylinder ($a_0=0.85$). We again consider an unconstrained cylinder (left column) and a constrained cylinder (right column), now for flow driven by an imposed pressure drop $\Delta{p}=0.025$. For clarity, we plot the results against the Lagrangian coordinate $R(r,t)=r-u_{s}$ on a linear horizontal scale. Unlike for the thick-walled cylinder (Fig.~\ref{fig:Thick}), the two different boundary conditions in this case result in strikingly different behaviour. For the unconstrained cylinder, the most important factors are the elasticity law and the treatment of the kinematics; the permeability law is relatively unimportant. For the constrained cylinder, all models exhibit nearly identical behaviour. \label{fig:Thin} }
\end{figure}

We now consider the other extreme geometry, a thin-walled cylinder, for a driving pressure drop of $\Delta{p}=0.025$ (Fig.~\ref{fig:Thin}). Note that this value of $\Delta{p}$ is more than one order of magnitude less than the value used for the thick-walled cylinder (Fig.~\ref{fig:Thick}). Despite this much smaller value of $\Delta{p}$, $\sigma_\theta^\prime$ here is comparable in magnitude to the thick-walled case while $u_s$ is much larger. We discuss these points in more detail in \S\ref{QL}.

For the unconstrained thin-walled cylinder (left column), $\phi_f$ (first row) is almost uniform across the domain, with a weak and roughly linear decrease from left to right. This behaviour is mirrored in $u_s$ (second row) and $\sigma_\theta^\prime$ (fourth row). The pressure also decreases roughly linearly from left to right, from $p(a,t)=\Delta{p}=0.025$ to $p(b,t)=0$, following classical linear poroelasticity for all models. Unlike for the thick-walled case, the permeability law is relatively unimportant for these quantities, whereas the kinematics and the elasticity law play much more prominent roles. Note that the kinematics consistently account for most of the difference between the L models and the N models (\textit{i.e.}, the Q models are closer to the N models than they are to the L models).

Unlike these other quantities, $\sigma_r^\prime$ does show a strong dependance on the permeability law. This suggests that the most direct impact of the permeability law is on $\sigma_r^\prime$, and this propagates to all other quantities when $\sigma_r^\prime$ is mechanically important (\textit{e.g.}, Fig.~\ref{fig:Thick}). For the unconstrained thin-walled cylinder, $\sigma_r^\prime$ vanishes at the boundaries and has an intermediate tensile maximum of order $10^{-3}$, whereas $\sigma_\theta^\prime$ is uniformly of order $10^{-1}$. As a result, the stark differences in $\sigma_r^\prime$ between the $k_0$ and $k_\mathrm{KC}$ models are ultimately unimportant.

\subsubsection{Constrained thin-walled cylinder}

For the same pressure drop, the constrained thin-walled cylinder exhibits strikingly different behaviour to the unconstrained thin-walled cylinder. Whereas the unconstrained cylinder expands almost uniformly by 20--70\%, the constrained cylinder is prevented from doing so. This results in much smaller displacements, with a maximum of order $10^{-3}$, making model choice essentially unimportant --- all models approach their asymptotic limit of classical linear poroelasticity (L-$k_0$). Note also that most of the material is in compression, with the porosity decreasing roughly linearly from a value just above $\phi_{f,0}$ at the inner boundary to a value noticeably below $\phi_{f,0}$ at the outer boundary. The displacement is weakly nonlinear, decreasing monotonically from left to right.

With regard to the flow rate $q$, we first note that the values of $q$ in this case are substantially larger than the corresponding values for the thick-walled cylinder despite the fact that $\Delta{p}$ is much smaller. To rationalise this, note that the relationship between $q$ and $a_0$ for a given $\Delta{p}$ is strongly nonlinear even for a rigid cylinder (\textit{i.e.}, Eq.~(\ref{q}) with $a=a_0$ and $b=b_0$). The same is also true for classical linear poroelasticity, where the same expression also applies. In other words, this difference in $q$ is due in large part to the fact that $a_0$ is much larger.

For the constrained thin-walled cylinder, $q$ is considerably smaller than for the unconstrained thin-walled cylinder (except for the L-$k_0$ case, where $q$ is independent of the boundary conditions). For the $k_0$ cases, this is because the cylinder expands substantially and almost uniformly, which decreases the ratio of $b$ to $a$ and increases the flow rate (see Eq.~(\ref{q})). This is true to a much lesser extent for the constrained cylinder since the displacements are much smaller. For the $k_\mathrm{KC}$ models, this increase in $q$ is substantially enhanced for the unconstrained cylinder by the noticeable increase in porosity and therefore permeability. The reverse occurs for the constrained cylinder, where the porosity decreases, leading a lower $q$ for the $k_\mathrm{KC}$ models than for the $k_0$ models. As for the thick-walled cylinder, both rigorous kinematics and nonlinear elasticity also lead to higher values of $q$ relative to their linearised counterparts. For the unconstrained cylinder, these effects are substantial; for the constrained cylinder, these effects are noticeably weaker than the impact of the permeability law. There is relatively little difference in $q$ across the six different models for the constrained cylinder, again because the displacements are necessarily small.

In this section, we have considered the implications of model choice in the context of two end-member geometries (thick-walled and thin-walled). We have shown that the error associated with linearisation depends strongly on factors such as geometry and boundary conditions. In the next section, we study the mechanics of the problem over the full transition from $a_0\ll{}1$ to $1-a_0\ll{}1$.

\subsection{Impact of geometry} \label{QL}

We now explore the parameter space more broadly, focusing on the importance of geometry ($a_0$) and driving ($q$ or $\Delta{p}$) while again fixing $\Gamma=0.4$ and $\phi_{f,0}=0.5$. Although the N-$k_\mathrm{KC}$ model is arguably the most `correct' of those considered above, it is much more computationally expensive than the other models. For simplicity, we restrict ourselves to the Q-$k_\mathrm{KC}$ model below. This model offers a good compromise between accuracy, robustness, and computational efficiency, demonstrating the same qualitative behaviour as the N-$k_\mathrm{KC}$ model for both end-member geometries and for both boundary conditions (see Appendix~\ref{supp:rheo}).

In Fig.~\ref{fig:geometry}, we consider the evolution of several key quantities as the inner radius $a_0$ varies continuously from $a_0 \ll 1$ (thick walls) to $1-a_0 \ll 1$ (thin walls). For a particular value of $a_0$, the flow can be driven by imposing either a fixed pressure drop $\Delta{p}$ or a fixed flow rate $q$; the other quantity ($q$ or $\Delta{p}$, respectively) is then calculated as part of the solution.\footnote{Note that one could instead impose both $\Delta{p}$ and $q$ and calculate $a_0$, which could be desirable in applications where $a_0$ is a design parameter to be used for targeting a particular combination of $\Delta{p}$ and $q$. We do not consider this case here.} We drive the flow with a fixed pressure drop $\Delta{p}$ and plot the results for several values of $\Delta{p}$ for unconstrained cylinders (left column) and constrained cylinders (right column). The resulting flow rate $q$ varies along these contours of fixed $\Delta{p}$ as shown in the last row.

Note that these same results can be presented in several different ways, which is useful for interpretation. Here, we show contours of fixed $\Delta{p}$ plotted against $a_0$ (Fig.~\ref{fig:geometry}). In Appendix~\ref{supp:geo}, we additionally show contours of fixed $q$ against $a_0$ (Fig.~\ref{fig:geo-a0_q}), contours of fixed $a_0$ against $q$ (Fig.~\ref{fig:geo-q_a0}), and contours of fixed $a_0$ against $\Delta{p}$ (Fig.~\ref{fig:geo-dp_a0}).

\begin{figure}[tb]
    \centering
    \includegraphics[width=5in]{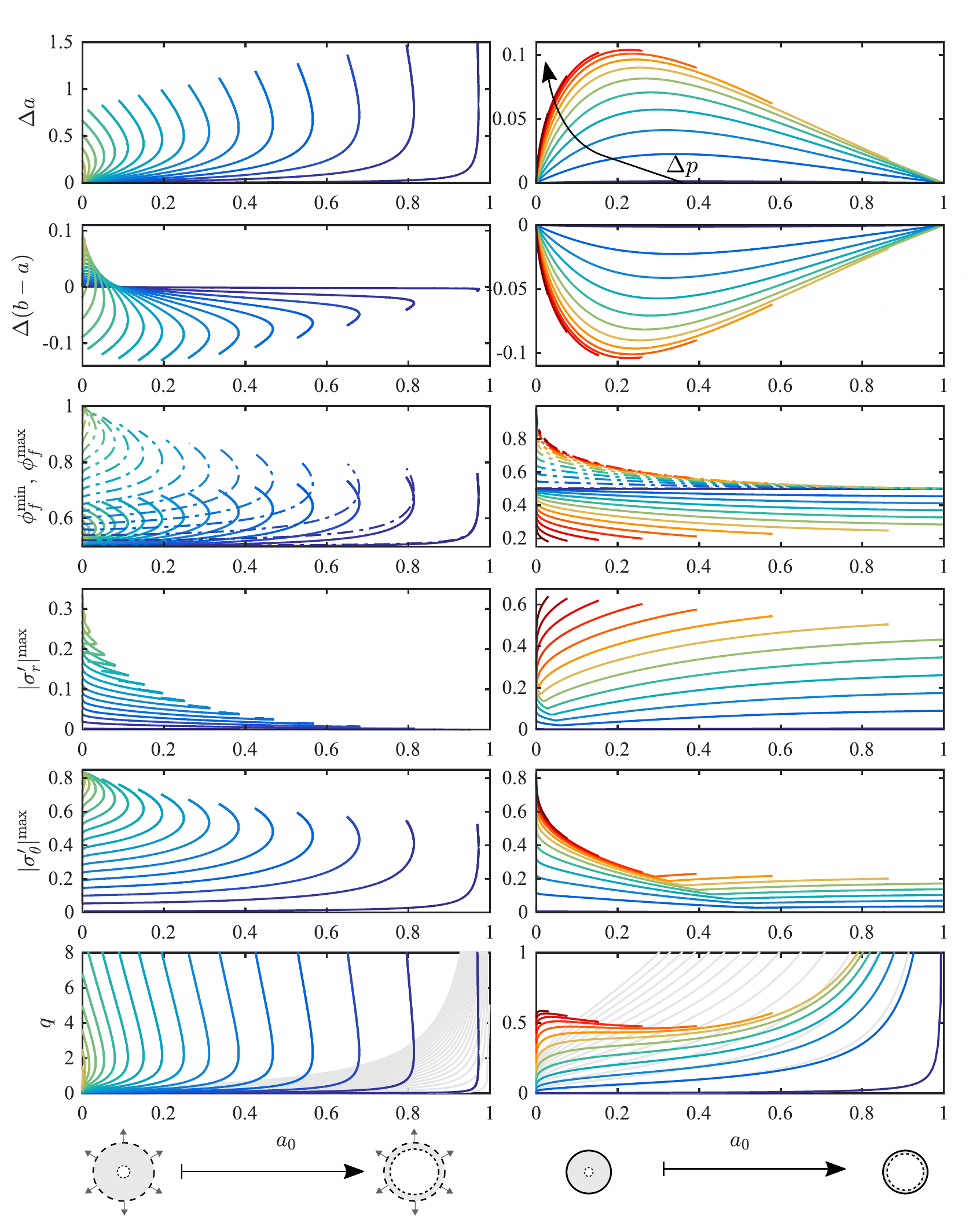}
    \caption{We explore the steady-state parameter space in more detail using the Q-$k_\mathrm{KC}$ model, plotting contours of fixed $\Delta{p}$ against $a_0$ for several key quantities for unconstrained cylinders (left, $\Delta{p} \in[0.005,0.5374]$, blue to yellow) and constrained cylinders (right, $ \Delta{p} \in [0.005,1.2]$, blue to red). We show the change in inner radius $\Delta{a}$ (first row); change in wall thickness $\Delta{(b-a)}$ (second row); minimum porosity $\phi_f^\mathrm{min}$ and maximum porosity $\phi_f^\mathrm{max}$ (solid and dot-dashed lines, respectively; third row); maximum absolute radial effective stress $|\sigma_r^\prime|^\mathrm{max}$ (fourth row) and maximum absolute azimuthal effective stress $|\sigma_\theta^\prime|^\mathrm{max}$ (fifth row); and flow rate $q$ (last row). We compare the latter with the reference flow rate $q_0$ that would occur for a rigid cylinder with the same initial geometry, $q_0=\Delta{p}\ln(b_0/a_0)^{-1}$ (grey lines). Note that the left and right columns use the same colour scale in $\Delta{p}$. \label{fig:geometry} }
\end{figure}

\subsection{Unconstrained cylinders}

For unconstrained cylinders (Fig.~\ref{fig:geometry}, left column), the most striking feature is the double-valued nature of all quantities for a certain range of $a_0$. Specifically, our results suggest that there exists a $\Delta{p}$-dependent maximum initial inner radius $a_0^\mathrm{max}(\Delta{p})$, above which the problem appears to have no solution and below which the problem appears to have two distinct solutions for at least some range of $a_0$. Although most of these contours terminate at some value of $a_0$ beyond which our numerical scheme is no longer able to converge to a solution, the existence of complete branches for larger values of $\Delta{p}$ suggests that all contours would continue smoothly back to $a_0=0$. For simplicity, we assume that this is indeed the case in the discussion below.

For $a_0>a_0^\mathrm{max}(\Delta{p})$, no steady-state solution exists. This suggests that, for a given value of $a_0$, there exists a maximum allowable driving pressure $\Delta{p}^\mathrm{max}(a_0)$ that can be supported (Fig.~\ref{fig:geo-dp_a0}). This maximum is an inherent feature of poromechanical coupling in a radial geometry. In the absence of a change in constitutive behaviour, applying a pressure drop larger than $\Delta{p}^\mathrm{max}(a_0)$ would lead to unbounded deformation and, ultimately, to material failure. The value of $\Delta{p}^\mathrm{max}$ is finite and positive for $0<a_0<1$, diverging as $a_0$ tends to zero and vanishing as $a_0$ tends to one.\footnote{The limit $a_0\to{}0$ corresponds to a line source in an infinite domain, for which no steady state exists. The limit $a_0\to{}1$ corresponds to vanishingly thin walls, which can support no load.} 

For $a_0<a_0^\mathrm{max}(\Delta{p})$, two distinct steady-state solutions exist for a given $a_0$. These correspond to a less-deformed solution and a more-deformed solution, where the latter is characterised by more extreme values of all quantities except for $|\sigma_r^\prime|^\mathrm{max}$. This implies that a given $\Delta{p}$ can lead to one of two different flow rates for the same cylinder: A lower flow rate in the less-deformed state or a higher flow rate in the more-deformed state. The classical balloon-inflation problem in nonlinear elasticity famously also exhibits multiple solutions in certain regions of its parameter space; in that case, the effect is purely kinematic and nonlinear-elastic. Here, this effect results from the nontrivial coupling of kinematics and poromechanics, even for a linear elasticity law. In the remainder of this section, we focus on the characteristics of these two solutions.

Flow drives all parts of the material radially outward ($u_r>0$ for all $r$), so that the inner and outer radii of the cylinder always increase, $a>a_0$ and $b>b_0$ (\textit{i.e.}, $\Delta{a}>0$, first row; $\Delta{b}>0$, not shown). The wall thickness $b-a$ may increase or decrease, depending on whether $\Delta{b}$ exceeds $\Delta{a}$ ($\Delta{(b-a)}$, second row). For $a_0\gtrsim{}0.1$, both solutions are characterised by a decrease in wall thickness. For $a_0\lesssim{}0.1$, the less-deformed solution instead corresponds to an increase in wall thickness. For $a_0\lesssim{}0.01$, both solutions correspond to an increase in wall thickness.

For all values of $a_0$ and $\Delta{p}$, both the minimum porosity $\phi_f^\mathrm{min}$ and the maximum porosity $\phi_f^\mathrm{max}$ exceed $\phi_{f,0}$ (third row; solid and dot-dashed lines, respectively). This implies that the porosity increases throughout the material ($\phi_f>\phi_{f,0}$ for all $r$), which further implies that the total cross-sectional area always increases, regardless of whether the wall thickness increases or decreases. For sufficiently small $\Delta{p}$, there exists a value of $a_0$ at which $\phi_f^\mathrm{min}$ and $\phi_f^\mathrm{max}$ intersect, implying the existence of a family of solutions with uniform porosity. The difference between $\phi_f^\mathrm{min}$ and $\phi_f^\mathrm{max}$ increases monotonically with $\Delta{p}$ such that this intersection no longer exists at high $\Delta{p}$ (Fig.~\ref{fig:geo-dp_a0}).

The maximum absolute azimuthal effective stress $|\sigma_\theta^\prime|^\mathrm{max}$ (fourth row) and the maximum absolute radial effective stress $|\sigma_r^\prime|^\mathrm{max}$ (fifth row) are relevant to material failure. The azimuthal component increases monotonically with $\Delta{p}$ along the less-deformed solution branch; the radial component exhibits a more complex behaviour, but $|\sigma_r^\prime|^\mathrm{max}<|\sigma_\theta^\prime|^\mathrm{max}$ for all $a_0$ and $\Delta{p}$ (Fig.~\ref{fig:geo-dp_a0}).

The flow rate $q$ exhibits the same striking feature as most other quantities---a region $a_0>a_0^\mathrm{max}(\Delta{p})$ characterised by no solution, and a region $a_0<a_0^\mathrm{max}(\Delta{p})$ characterised by two solutions (last row; coloured lines). We compare the actual flow rate $q$ with the reference flow rate $q_0$ that would occur for the same $\Delta{p}$ for a rigid cylinder with the same initial geometry, $q_0=\Delta{p}\ln(b_0/a_0)^{-1}$ (last row; grey lines). This reference flow rate is equivalent to the prediction of classical linear poroelasticity (L-$k_0$), and it diverges for all $\Delta{p}$ as $a_0$ tends to one. Note that $q>q_0$ for all $a_0$ and $\Delta{p}$---that is, a deformable unconfined cylinder will always conduct a higher flow rate than a rigid cylinder of the same initial geometry, and this is a nonlinear effect.

\subsection{Constrained cylinders}

Constrained cylinders exhibit qualitatively different behaviour (Fig.~\ref{fig:geometry}, right column) --- a single solution exists for all values of $a_0$, and all quantities vary monotonically with $\Delta{p}$. Note that we expect unconstrained and constrained cylinders to approach the same limiting behaviour for $a_0 \ll 1$, as noted above in the context of Fig.~\ref{fig:Thick}.

The change in inner radius $\Delta{a}$ is strictly positive, tending to zero for both small $a_0$ and large $a_0$. In the former limit, this is because $\Delta{a}$ decreases with $a_0$ for fixed $\Delta{p}$; in the latter limit, this is because $b$ is fixed and the material has nowhere to go. The change in wall thickness is equal and opposite to the change in inner radius, $\Delta{(b-a)}=-\Delta{a}$, and is therefore strictly negative. That is, the walls always get thinner. As a result, the cross-sectional area always decreases and the average porosity (and thus $\phi_f^\mathrm{min}$) must always be less than $\phi_{f,0}$. However, $\phi_f^\mathrm{max}$ is still always greater than $\phi_{f,0}$. The difference between $\phi_f^\mathrm{max}$ and $\phi_f^\mathrm{min}$ increases with $\Delta{p}$ (Fig.~\ref{fig:geo-dp_a0}) and is roughly constant with $a_0$. For a thin-walled cylinder, $\phi_f^\mathrm{max}$ is close to $\phi_{f,0}$ while $\phi_f^\mathrm{min}$ is substantially below $\phi_{f,0}$. For a thick-walled cylinder, $\phi_f^\mathrm{min}$ is close to $\phi_{f,0}$ while $\phi_f^\mathrm{max}$ is substantially above $\phi_{f,0}$. Note that the latter scenario respects the constraint on the average porosity by virtue of the fact that the large porosities are localised to a small region near the inner radius while the rest of the cylinder (the vast majority) is weakly compressed. The azimuthal stress $|\sigma_\theta^\prime|^\mathrm{max}$ decreases with $a_0$ for small $a_0$ and increases gently with $a_0$ for large $a_0$, tending to a finite, nonzero value as $a_0$ tends to one. The radial stress $|\sigma_r^\prime|^\mathrm{max}$ exhibits a similar trend, with the transition from decreasing to increasing occurring at a much smaller value of $a_0$. For both stress components, this transition occurs at a corner that corresponds to a transition in the maximum absolute value of the stress from tensile near/at the inner radius (radial/azimuthal) to compressive at the outer radius (both).

The flow rate $q$ is weakly non-monotonic in $a_0$ for small $a_0$ and large $\Delta{p}$, implying that two different values of $a_0$ can lead to the same combination of $\Delta{p}$ and $q$. Comparing the actual flow rate $q$ to the reference flow rate $q_0$ (rigid cylinder or L-$k_0$ model, grey lines), we find that a constrained deformable cylinder will conduct a larger flow rate than a rigid cylinder if the walls are thick, but a smaller flow rate than a rigid one if the walls are thin; this is in contrast to an unconstrained deformable cylinder, which always conducts a larger flow rate than a rigid one. This effect is amplified as $\Delta{p}$ increases, but its magnitude is relatively modest; $q$ decreases from a few tens of percent above $q_0$ to a few tens of percent below $q_0$ over the full range of $a_0$. For an unconstrained cylinder, in contrast, deformation dominates the flow rate as $a_0$ approaches $a_0^\mathrm{max}$.

\subsection{Force balance}

\begin{figure}[tb]
    \centering
    \includegraphics[width=5in]{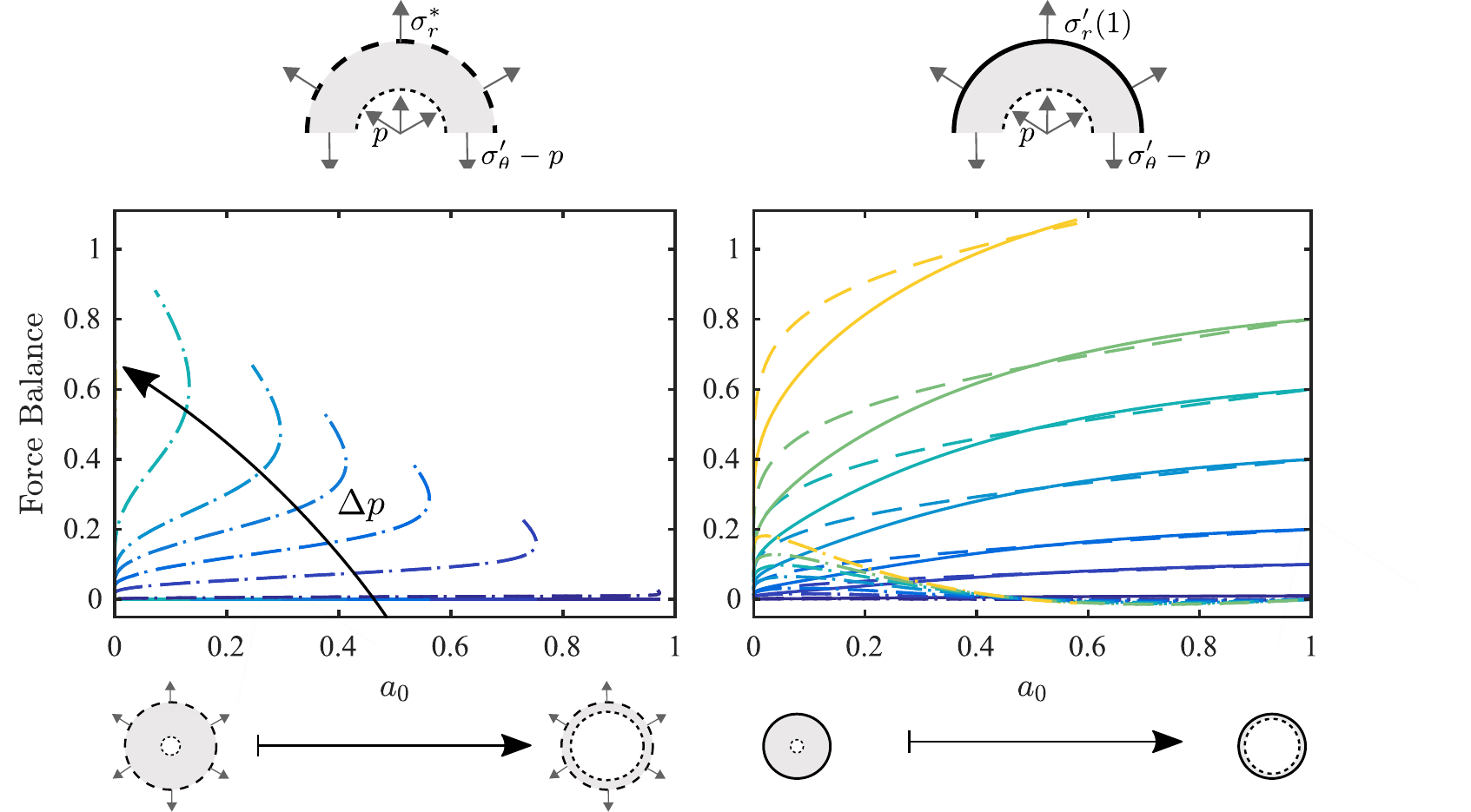}
    \caption{Flow leads to a net pressure force $F_p$ (dashed lines) that must be supported by a combination of force due to internal azimuthal stress $F_\theta$ (dot-dashed lines) and force due to external radial traction $F_r$ (solid lines). We plot these forces for unconstrained cylinders for $\Delta{p}\in[0.005,0.5]$ (left), and for constrained cylinders for $\Delta{p}\in[0.005,0.6]$ (right). The colour scale for $\Delta{p}$ is the same as in Fig.~\ref{fig:geometry}. For unconstrained cylinders, note that $F_r\equiv{}0$ and $F_\theta\equiv{}F_p$. \label{fig:geo_force} }
\end{figure}

Flow always forces the material radially outward. This loading must be supported through a combination of internal azimuthal stress and external radial traction. To investigate these mechanics in more detail, we consider a macroscopic balance of the `vertical' components of the forces acting on one-half of the annular cross-section of the cylinder (see diagrams, top of Fig.~\ref{fig:geo_force}). The `vertical' components of the forces due to fluid or pore-pressure loading $F_p$, internal azimuthal stress $F_\theta$, and external radial traction $F_r$ are given by
\begin{equation}\label{eq:Forces}
    F_p =2a\Delta{p} +2\int^b_a p\,\mathrm{d}r \,,\quad F_\theta=2\int^b_a \sigma_\theta^\prime\,\mathrm{d}r \,,\quad \mathrm{and}\quad F_r=-2b\sigma_r^\prime(b),
\end{equation} 
and macroscopic force balance requires that $F_p = F_\theta + F_r$. We plot these quantities in Fig.~\ref{fig:geo_force} for unconstrained cylinders (left column) and constrained cylinders (right column). Note that, as with Fig.~\ref{fig:geometry}, these results can be presented in several different ways (see Fig.~\ref{fig:ForceBalanceCombo}).

For unconstrained cylinders, $F_r\equiv{}0$ and therefore $F_\theta\equiv{}F_p$. These two nontrivial force components increase as $\Delta{p}$ increases along the less-deformed solution branch. These quantities ultimately mirror the behaviour shown in Fig.~\ref{fig:geometry}---two solutions exist for $a_0<a_0^\mathrm{max}(\Delta{p})$, one corresponding to less deformation and smaller forces and the other corresponding to more deformation and larger forces.

For constrained cylinders, $F_r$ will be determined implicitly to satisfy the condition that $u_s(1)=0$. For fixed $\Delta{p}$, both $F_p$ and $F_r$ increase monotonically with $a_0$. For $a_0\lesssim{}0.05$, $F_\theta$ is similar in magnitude to $F_r$ and increases with $a_0$; for $a_0\gtrsim{}0.05$, however, $F_\theta$ decreases rapidly with $a_0$ and ultimately becomes weakly negative but negligible in the overall force balance. In other words, the outer boundary supports most of the fluid loading for a cylinder with moderate to thin walls. Note that $F_r<F_p$ for $a_0\lesssim{}0.5$ since $F_\theta>0$, but $F_r>F_p$ for $a_0\gtrsim{}0.5$ since $F_\theta<0$.

\section{Conclusion}

Despite being a classical topic in geomechanics and in biophysics, radial poroelastic deformation has not previously been systematically explored, particularly in the context of large deformations. To assess the qualitative and quantitative impacts of large deformations, we considered six different models in the context of two end-member geometries (thick-walled and thin-walled) and two different outer boundary conditions (unconstrained and constrained). We showed that the impacts of nonlinear kinematics, nonlinear elasticity, and deformation-dependent permeability depend strongly on geometry and boundary conditions, as does the relative importance of these facets of nonlinearity. For example, the mechanical response of an unconstrained thin-walled cylinder to an imposed pressure drop is dominated by kinematics and elasticity, although the permeability law exerts a strong control on the resulting flow rate through the material; for the same pressure drop, a constrained thin-walled cylinder is limited to much smaller deformations and exhibits what is essentially a linear-poroelastic response (Fig.~\ref{fig:Thin}). In contrast, the mechanical response of a thick-walled cylinder is much less sensitive to constraint, although the outer boundary condition has a strong impact on the flow rate when the permeability is deformation-dependent~(Fig.~\ref{fig:Thick}).

To explore the importance of geometry and constraint in more detail, we then focused on a model that includes rigorous nonlinear kinematics and deformation-dependent permeability, but with the simplification of linear elasticity (Q-$k_\mathrm{KC}$). This model captures the qualitative and quantitative impacts of large deformations (see Fig.~F1), but is more computationally convenient than a fully nonlinear model. We showed that, for an unconstrained cylinder, a given initial inner radius can conduct an arbitrarily large flow rate but can only support a finite maximum pressure drop, and this maximum allowable pressure drop increases with the thickness of the walls (Figs.~\ref{fig:geometry}, \ref{fig:geo_force}, and E3). For a pressure drop less than this maximum, our results suggest that two valid solutions exist---a less-deformed state with a lower flow rate and a more-deformed state with a higher flow rate. A constrained cylinder, in contrast, can support an arbitrarily large pressure drop but can only conduct a finite maximum flow rate, and this maximum flow rate is nonmonotonic in the wall thickness (see Figs.~\ref{fig:geo-a0_q}--\ref{fig:geo-dp_a0}). These behaviours are mirrored in the corresponding force balances (Figs.~\ref{fig:geo_force} and \ref{fig:ForceBalanceCombo}).

We have assumed here that the constitutive response of the solid skeleton remains elastic for arbitrarily large deformations. This is relevant to biomedical applications such as fluid permeation through artery walls, and to the design of radial filters. In geomechanical applications, however, large deformations are typically the result of material failure through plasticity or fracture, which will lead to a fundamentally different constitutive behaviour in the solid. Additionally, it may be relevant for many biomedical and geophysical applications to couple the poroelastic domain to different surface phenomena, such as free external flows. These behaviours may be the subject of future work.

We conclude by noting that, in addition to providing fundamental physical insight, our results and numerical codes could serve as a useful benchmark for general numerical-simulation tools (\textit{e.g.,} finite-element codes). Relatively few benchmarks are available in the context of large-deformation poroelasticity.



The authors are grateful to EPSRC for support in the form of a Doctoral Training Award to LCA. The authors also thank Simon~Mathias for helpful discussions and advice related to the Chebyshev pseudospectral method.

\clearpage

\begin{thebibliography}{37}%
\makeatletter
\providecommand \@ifxundefined [1]{%
 \@ifx{#1\undefined}
}%
\providecommand \@ifnum [1]{%
 \ifnum #1\expandafter \@firstoftwo
 \else \expandafter \@secondoftwo
 \fi
}%
\providecommand \@ifx [1]{%
 \ifx #1\expandafter \@firstoftwo
 \else \expandafter \@secondoftwo
 \fi
}%
\providecommand \natexlab [1]{#1}%
\providecommand \enquote  [1]{``#1''}%
\providecommand \bibnamefont  [1]{#1}%
\providecommand \bibfnamefont [1]{#1}%
\providecommand \citenamefont [1]{#1}%
\providecommand \href@noop [0]{\@secondoftwo}%
\providecommand \href [0]{\begingroup \@sanitize@url \@href}%
\providecommand \@href[1]{\@@startlink{#1}\@@href}%
\providecommand \@@href[1]{\endgroup#1\@@endlink}%
\providecommand \@sanitize@url [0]{\catcode `\\12\catcode `\$12\catcode
  `\&12\catcode `\#12\catcode `\^12\catcode `\_12\catcode `\%12\relax}%
\providecommand \@@startlink[1]{}%
\providecommand \@@endlink[0]{}%
\providecommand \url  [0]{\begingroup\@sanitize@url \@url }%
\providecommand \@url [1]{\endgroup\@href {#1}{\urlprefix }}%
\providecommand \urlprefix  [0]{URL }%
\providecommand \Eprint [0]{\href }%
\providecommand \doibase [0]{http://dx.doi.org/}%
\providecommand \selectlanguage [0]{\@gobble}%
\providecommand \bibinfo  [0]{\@secondoftwo}%
\providecommand \bibfield  [0]{\@secondoftwo}%
\providecommand \translation [1]{[#1]}%
\providecommand \BibitemOpen [0]{}%
\providecommand \bibitemStop [0]{}%
\providecommand \bibitemNoStop [0]{.\EOS\space}%
\providecommand \EOS [0]{\spacefactor3000\relax}%
\providecommand \BibitemShut  [1]{\csname bibitem#1\endcsname}%
\let\auto@bib@innerbib\@empty
\bibitem [{\citenamefont {Randolph}\ and\ \citenamefont
  {Wroth}(1979)}]{randolph1979analytical}%
  \BibitemOpen
  \bibfield  {author} {\bibinfo {author} {\bibfnamefont {M.~F.}\ \bibnamefont
  {Randolph}}\ and\ \bibinfo {author} {\bibfnamefont {C.~P.}\ \bibnamefont
  {Wroth}},\ }\bibfield  {title} {\enquote {\bibinfo {title} {An analytical
  solution for the consolidation around a driven pile},}\ }\href@noop {}
  {\bibfield  {journal} {\bibinfo  {journal} {International Journal for
  Numerical and Analytical Methods in Geomechanics}\ }\textbf {\bibinfo
  {volume} {3}},\ \bibinfo {pages} {217--229} (\bibinfo {year}
  {1979})}\BibitemShut {NoStop}%
\bibitem [{\citenamefont {Seth}\ and\ \citenamefont
  {Gray}(1968)}]{seth1968transientb}%
  \BibitemOpen
  \bibfield  {author} {\bibinfo {author} {\bibfnamefont {M.~S.}\ \bibnamefont
  {Seth}}\ and\ \bibinfo {author} {\bibfnamefont {K.~E.}\ \bibnamefont
  {Gray}},\ }\bibfield  {title} {\enquote {\bibinfo {title} {Transient stresses
  and displacement around a wellbore due to fluid flow in transversely
  isotropic, porous media: {II}. {F}inite reservoirs},}\ }\href@noop {}
  {\bibfield  {journal} {\bibinfo  {journal} {Society of Petroleum Engineers
  Journal}\ }\textbf {\bibinfo {volume} {8}},\ \bibinfo {pages} {79--86}
  (\bibinfo {year} {1968})}\BibitemShut {NoStop}%
\bibitem [{\citenamefont {Rice}\ and\ \citenamefont
  {Clearly}(1976)}]{rice1976some}%
  \BibitemOpen
  \bibfield  {author} {\bibinfo {author} {\bibfnamefont {J.~R.}\ \bibnamefont
  {Rice}}\ and\ \bibinfo {author} {\bibfnamefont {M.~P.}\ \bibnamefont
  {Clearly}},\ }\bibfield  {title} {\enquote {\bibinfo {title} {Some basic
  stress diffusion solutions for fluid-saturated elastic porous media with
  compressible constituents},}\ }\href@noop {} {\bibfield  {journal} {\bibinfo
  {journal} {Reviews of Geophysics and Space Physics}\ }\textbf {\bibinfo
  {volume} {14}} (\bibinfo {year} {1976})}\BibitemShut {NoStop}%
\bibitem [{\citenamefont {Detournay}\ and\ \citenamefont
  {Cheng}(1988)}]{detournay1988poroelastic}%
  \BibitemOpen
  \bibfield  {author} {\bibinfo {author} {\bibfnamefont {E.}~\bibnamefont
  {Detournay}}\ and\ \bibinfo {author} {\bibfnamefont {A.~H.~D.}\ \bibnamefont
  {Cheng}},\ }\bibfield  {title} {\enquote {\bibinfo {title} {Poroelastic
  response of a borehole in a non-hydrostatic stress field},}\ }\href@noop {}
  {\bibfield  {journal} {\bibinfo  {journal} {International Journal of Rock
  Mechanics and Mining Sciences \& Geomechanics Abstracts}\ }\textbf {\bibinfo
  {volume} {25}},\ \bibinfo {pages} {171--182} (\bibinfo {year}
  {1988})}\BibitemShut {NoStop}%
\bibitem [{\citenamefont {Sciarra}\ \emph {et~al.}(2005)\citenamefont
  {Sciarra}, \citenamefont {{dell'Isola}},\ and\ \citenamefont
  {Hutter}}]{sciarra-intjgeomech-2005}%
  \BibitemOpen
  \bibfield  {author} {\bibinfo {author} {\bibfnamefont {G.}~\bibnamefont
  {Sciarra}}, \bibinfo {author} {\bibfnamefont {F.}~\bibnamefont
  {{dell'Isola}}}, \ and\ \bibinfo {author} {\bibfnamefont {K.}~\bibnamefont
  {Hutter}},\ }\bibfield  {title} {\enquote {\bibinfo {title} {Dilatational and
  compacting behavior around a cylindrical cavern leached out in a solid-fluid
  elastic rock salt},}\ }\href {\doibase 10.1061/(ASCE)1532-3641(2005)5:3(233)}
  {\bibfield  {journal} {\bibinfo  {journal} {International Journal of
  Geomechanics}\ }\textbf {\bibinfo {volume} {5}},\ \bibinfo {pages} {233--243}
  (\bibinfo {year} {2005})}\BibitemShut {NoStop}%
\bibitem [{\citenamefont {van Gerwen}\ \emph {et~al.}(2012)\citenamefont {van
  Gerwen}, \citenamefont {Dankelman},\ and\ \citenamefont {van~den
  Dobbelsteen}}]{van2012needle}%
  \BibitemOpen
  \bibfield  {author} {\bibinfo {author} {\bibfnamefont {D.~J.}\ \bibnamefont
  {van Gerwen}}, \bibinfo {author} {\bibfnamefont {J.}~\bibnamefont
  {Dankelman}}, \ and\ \bibinfo {author} {\bibfnamefont {J.~J.}\ \bibnamefont
  {van~den Dobbelsteen}},\ }\bibfield  {title} {\enquote {\bibinfo {title}
  {Needle--tissue interaction forces---a survey of experimental data},}\
  }\href@noop {} {\bibfield  {journal} {\bibinfo  {journal} {Medical
  Engineering \& Physics}\ }\textbf {\bibinfo {volume} {34}},\ \bibinfo {pages}
  {665--680} (\bibinfo {year} {2012})}\BibitemShut {NoStop}%
\bibitem [{\citenamefont {Kenyon}(1979)}]{kenyon1979mathematical}%
  \BibitemOpen
  \bibfield  {author} {\bibinfo {author} {\bibfnamefont {D.~E.}\ \bibnamefont
  {Kenyon}},\ }\bibfield  {title} {\enquote {\bibinfo {title} {A mathematical
  model of water flux through aortic tissue},}\ }\href@noop {} {\bibfield
  {journal} {\bibinfo  {journal} {Bulletin of Mathematical Biology}\ }\textbf
  {\bibinfo {volume} {41}},\ \bibinfo {pages} {79--90} (\bibinfo {year}
  {1979})}\BibitemShut {NoStop}%
\bibitem [{\citenamefont {Jayaraman}(1983)}]{jayaraman1983water}%
  \BibitemOpen
  \bibfield  {author} {\bibinfo {author} {\bibfnamefont {G.}~\bibnamefont
  {Jayaraman}},\ }\bibfield  {title} {\enquote {\bibinfo {title} {Water
  transport in the arterial wall---{A} theoretical study},}\ }\href@noop {}
  {\bibfield  {journal} {\bibinfo  {journal} {Journal of biomechanics}\
  }\textbf {\bibinfo {volume} {16}},\ \bibinfo {pages} {833--840} (\bibinfo
  {year} {1983})}\BibitemShut {NoStop}%
\bibitem [{\citenamefont {Klanchar}\ and\ \citenamefont
  {Tarbell}(1987)}]{klanchar1987modeling}%
  \BibitemOpen
  \bibfield  {author} {\bibinfo {author} {\bibfnamefont {M.}~\bibnamefont
  {Klanchar}}\ and\ \bibinfo {author} {\bibfnamefont {J.~M.}\ \bibnamefont
  {Tarbell}},\ }\bibfield  {title} {\enquote {\bibinfo {title} {Modeling water
  flow through arterial tissue},}\ }\href@noop {} {\bibfield  {journal}
  {\bibinfo  {journal} {Bulletin of Mathematical Biology}\ }\textbf {\bibinfo
  {volume} {49}},\ \bibinfo {pages} {651--669} (\bibinfo {year}
  {1987})}\BibitemShut {NoStop}%
\bibitem [{\citenamefont {Barry}\ and\ \citenamefont
  {Aldis}(1993)}]{barry1993radial}%
  \BibitemOpen
  \bibfield  {author} {\bibinfo {author} {\bibfnamefont {S.~I.}\ \bibnamefont
  {Barry}}\ and\ \bibinfo {author} {\bibfnamefont {G.K.}\ \bibnamefont
  {Aldis}},\ }\bibfield  {title} {\enquote {\bibinfo {title} {Radial flow
  through deformable porous shells},}\ }\href@noop {} {\bibfield  {journal}
  {\bibinfo  {journal} {The Journal of the Australian Mathematical Society.
  Series B. Applied Mathematics}\ }\textbf {\bibinfo {volume} {34}},\ \bibinfo
  {pages} {333--354} (\bibinfo {year} {1993})}\BibitemShut {NoStop}%
\bibitem [{\citenamefont {Barry}\ and\ \citenamefont
  {Mercer}(1998)}]{barry1998effect}%
  \BibitemOpen
  \bibfield  {author} {\bibinfo {author} {\bibfnamefont {S.~I.}\ \bibnamefont
  {Barry}}\ and\ \bibinfo {author} {\bibfnamefont {G.~N.}\ \bibnamefont
  {Mercer}},\ }\bibfield  {title} {\enquote {\bibinfo {title} {Effect of a
  moving boundary on the deformation of a poro-elastic cylinder},}\ }\href@noop
  {} {\bibfield  {journal} {\bibinfo  {journal} {ANZIAM Journal}\ }\textbf
  {\bibinfo {volume} {39}},\ \bibinfo {pages} {627--666} (\bibinfo {year}
  {1998})}\BibitemShut {NoStop}%
\bibitem [{\citenamefont {Skotheim}\ and\ \citenamefont
  {Mahadevan}(2005)}]{skotheim2005physical}%
  \BibitemOpen
  \bibfield  {author} {\bibinfo {author} {\bibfnamefont {Jan~M}\ \bibnamefont
  {Skotheim}}\ and\ \bibinfo {author} {\bibfnamefont {L}~\bibnamefont
  {Mahadevan}},\ }\bibfield  {title} {\enquote {\bibinfo {title} {Physical
  limits and design principles for plant and fungal movements},}\ }\href@noop
  {} {\bibfield  {journal} {\bibinfo  {journal} {Science}\ }\textbf {\bibinfo
  {volume} {308}},\ \bibinfo {pages} {1308--1310} (\bibinfo {year}
  {2005})}\BibitemShut {NoStop}%
\bibitem [{\citenamefont {Reichold}\ \emph {et~al.}(2009)\citenamefont
  {Reichold}, \citenamefont {Stampanoni}, \citenamefont {Keller}, \citenamefont
  {Buck}, \citenamefont {Jenny},\ and\ \citenamefont
  {Weber}}]{reichold2009vascular}%
  \BibitemOpen
  \bibfield  {author} {\bibinfo {author} {\bibfnamefont {Johannes}\
  \bibnamefont {Reichold}}, \bibinfo {author} {\bibfnamefont {Marco}\
  \bibnamefont {Stampanoni}}, \bibinfo {author} {\bibfnamefont {Anna~Lena}\
  \bibnamefont {Keller}}, \bibinfo {author} {\bibfnamefont {Alfred}\
  \bibnamefont {Buck}}, \bibinfo {author} {\bibfnamefont {Patrick}\
  \bibnamefont {Jenny}}, \ and\ \bibinfo {author} {\bibfnamefont {Bruno}\
  \bibnamefont {Weber}},\ }\bibfield  {title} {\enquote {\bibinfo {title}
  {Vascular graph model to simulate the cerebral blood flow in realistic
  vascular networks},}\ }\href@noop {} {\bibfield  {journal} {\bibinfo
  {journal} {Journal of Cerebral Blood Flow \& Metabolism}\ }\textbf {\bibinfo
  {volume} {29}},\ \bibinfo {pages} {1429--1443} (\bibinfo {year}
  {2009})}\BibitemShut {NoStop}%
\bibitem [{\citenamefont {Chou}\ \emph {et~al.}(2013)\citenamefont {Chou},
  \citenamefont {Wang},\ and\ \citenamefont {Fane}}]{chou2013robust}%
  \BibitemOpen
  \bibfield  {author} {\bibinfo {author} {\bibfnamefont {Shuren}\ \bibnamefont
  {Chou}}, \bibinfo {author} {\bibfnamefont {Rong}\ \bibnamefont {Wang}}, \
  and\ \bibinfo {author} {\bibfnamefont {Anthony~G}\ \bibnamefont {Fane}},\
  }\bibfield  {title} {\enquote {\bibinfo {title} {Robust and high performance
  hollow fiber membranes for energy harvesting from salinity gradients by
  pressure retarded osmosis},}\ }\href@noop {} {\bibfield  {journal} {\bibinfo
  {journal} {Journal of Membrane Science}\ }\textbf {\bibinfo {volume} {448}},\
  \bibinfo {pages} {44--54} (\bibinfo {year} {2013})}\BibitemShut {NoStop}%
\bibitem [{\citenamefont {Biot}(1941)}]{biot1941general}%
  \BibitemOpen
  \bibfield  {author} {\bibinfo {author} {\bibfnamefont {Maurice~A}\
  \bibnamefont {Biot}},\ }\bibfield  {title} {\enquote {\bibinfo {title}
  {General theory of three-dimensional consolidation},}\ }\href@noop {}
  {\bibfield  {journal} {\bibinfo  {journal} {Journal of applied physics}\
  }\textbf {\bibinfo {volume} {12}},\ \bibinfo {pages} {155--164} (\bibinfo
  {year} {1941})}\BibitemShut {NoStop}%
\bibitem [{\citenamefont {Wang}(2000)}]{wang2000theory}%
  \BibitemOpen
  \bibfield  {author} {\bibinfo {author} {\bibfnamefont {Herbert}\ \bibnamefont
  {Wang}},\ }\href@noop {} {\emph {\bibinfo {title} {Theory of linear
  poroelasticity with applications to geomechanics and hydrogeology}}}\
  (\bibinfo  {publisher} {Princeton University Press},\ \bibinfo {year}
  {2000})\BibitemShut {NoStop}%
\bibitem [{\citenamefont {Argoubi}\ and\ \citenamefont
  {Shirazi-Adl}(1996)}]{argoubi-jbiomech-1996}%
  \BibitemOpen
  \bibfield  {author} {\bibinfo {author} {\bibfnamefont {M}~\bibnamefont
  {Argoubi}}\ and\ \bibinfo {author} {\bibfnamefont {A}~\bibnamefont
  {Shirazi-Adl}},\ }\bibfield  {title} {\enquote {\bibinfo {title} {Poroelastic
  creep response analysis of a lumbar motion segment in compression},}\ }\href
  {\doibase 10.1016/0021-9290(96)00035-8} {\bibfield  {journal} {\bibinfo
  {journal} {Journal of Biomechanics}\ }\textbf {\bibinfo {volume} {29}},\
  \bibinfo {pages} {1331--1339} (\bibinfo {year} {1996})}\BibitemShut {NoStop}%
\bibitem [{\citenamefont {Federico}\ and\ \citenamefont
  {Grillo}(2012)}]{federico-mechmater-2012}%
  \BibitemOpen
  \bibfield  {author} {\bibinfo {author} {\bibfnamefont {S.}~\bibnamefont
  {Federico}}\ and\ \bibinfo {author} {\bibfnamefont {A.}~\bibnamefont
  {Grillo}},\ }\bibfield  {title} {\enquote {\bibinfo {title} {Elasticity and
  permeability of porous fibre-reinforced materials under large
  deformations},}\ }\href {\doibase 10.1016/j.mechmat.2011.07.010} {\bibfield
  {journal} {\bibinfo  {journal} {Mechanics of Materials}\ }\textbf {\bibinfo
  {volume} {44}},\ \bibinfo {pages} {58--71} (\bibinfo {year}
  {2012})}\BibitemShut {NoStop}%
\bibitem [{\citenamefont {Tomic}\ \emph {et~al.}(2014)\citenamefont {Tomic},
  \citenamefont {Grillo},\ and\ \citenamefont
  {Federico}}]{tomic-imajapplmath-2014}%
  \BibitemOpen
  \bibfield  {author} {\bibinfo {author} {\bibfnamefont {A.}~\bibnamefont
  {Tomic}}, \bibinfo {author} {\bibfnamefont {A.}~\bibnamefont {Grillo}}, \
  and\ \bibinfo {author} {\bibfnamefont {S.}~\bibnamefont {Federico}},\
  }\bibfield  {title} {\enquote {\bibinfo {title} {Poroelastic materials
  reinforced by statistically oriented fibers---numerical implementation and
  applications to articular cartilage},}\ }\href {\doibase
  10.1093/imamat/hxu039} {\bibfield  {journal} {\bibinfo  {journal} {{IMA}
  Journal of Applied Mathematics}\ }\textbf {\bibinfo {volume} {79}},\ \bibinfo
  {pages} {1027--1059} (\bibinfo {year} {2014})}\BibitemShut {NoStop}%
\bibitem [{\citenamefont {Vuong}\ \emph {et~al.}(2015)\citenamefont {Vuong},
  \citenamefont {Yoshihara},\ and\ \citenamefont
  {Wall}}]{vuong-compmethapplmecheng-2015}%
  \BibitemOpen
  \bibfield  {author} {\bibinfo {author} {\bibfnamefont {A.-T.}\ \bibnamefont
  {Vuong}}, \bibinfo {author} {\bibfnamefont {L.}~\bibnamefont {Yoshihara}}, \
  and\ \bibinfo {author} {\bibfnamefont {W.~A.}\ \bibnamefont {Wall}},\
  }\bibfield  {title} {\enquote {\bibinfo {title} {A general approach for
  modeling interacting flow through porous media under finite deformations},}\
  }\href {\doibase 10.1016/j.cma.2014.08.018} {\bibfield  {journal} {\bibinfo
  {journal} {Computer Methods in Applied Mechanics and Engineering}\ }\textbf
  {\bibinfo {volume} {283}},\ \bibinfo {pages} {1240--1259} (\bibinfo {year}
  {2015})}\BibitemShut {NoStop}%
\bibitem [{\citenamefont {Uzuoka}\ and\ \citenamefont
  {Borja}(2011)}]{uzuoka-intjnag-2011}%
  \BibitemOpen
  \bibfield  {author} {\bibinfo {author} {\bibfnamefont {R.}~\bibnamefont
  {Uzuoka}}\ and\ \bibinfo {author} {\bibfnamefont {R.~I.}\ \bibnamefont
  {Borja}},\ }\bibfield  {title} {\enquote {\bibinfo {title} {Dynamics of
  unsaturated poroelastic solids at finite strain},}\ }\href {\doibase
  10.1002/nag.1061} {\bibfield  {journal} {\bibinfo  {journal} {International
  Journal for Numerical and Analytical Methods in Geomechanics}\ }\textbf
  {\bibinfo {volume} {36}},\ \bibinfo {pages} {1535--1573} (\bibinfo {year}
  {2011})}\BibitemShut {NoStop}%
\bibitem [{\citenamefont {Song}\ and\ \citenamefont
  {Borja}(2014)}]{song-vzj-2014}%
  \BibitemOpen
  \bibfield  {author} {\bibinfo {author} {\bibfnamefont {X.}~\bibnamefont
  {Song}}\ and\ \bibinfo {author} {\bibfnamefont {R.~I.}\ \bibnamefont
  {Borja}},\ }\bibfield  {title} {\enquote {\bibinfo {title} {Finite
  deformation and fluid flow in unsaturated soils with random heterogeneity},}\
  }\href {\doibase 10.2136/vzj2013.07.0131} {\bibfield  {journal} {\bibinfo
  {journal} {Vadose Zone Journal}\ }\textbf {\bibinfo {volume} {13}} (\bibinfo
  {year} {2014}),\ 10.2136/vzj2013.07.0131}\BibitemShut {NoStop}%
\bibitem [{\citenamefont {Borja}\ and\ \citenamefont
  {Choo}(2016)}]{borja-compmethapplmecheng-2016}%
  \BibitemOpen
  \bibfield  {author} {\bibinfo {author} {\bibfnamefont {R.~I.}\ \bibnamefont
  {Borja}}\ and\ \bibinfo {author} {\bibfnamefont {J.}~\bibnamefont {Choo}},\
  }\bibfield  {title} {\enquote {\bibinfo {title} {{Cam-Clay} plasticity, {Part
  VIII}: {A} constitutive framework for porous materials with evolving internal
  structure},}\ }\href {\doibase 10.1016/j.cma.2016.06.016} {\bibfield
  {journal} {\bibinfo  {journal} {Computer Methods in Applied Mechanics and
  Engineering}\ }\textbf {\bibinfo {volume} {309}},\ \bibinfo {pages}
  {653--679} (\bibinfo {year} {2016})}\BibitemShut {NoStop}%
\bibitem [{\citenamefont {Beavers}\ \emph {et~al.}(1981)\citenamefont
  {Beavers}, \citenamefont {Wittenberg},\ and\ \citenamefont
  {Sparrow}}]{beavers1981fluidpart2}%
  \BibitemOpen
  \bibfield  {author} {\bibinfo {author} {\bibfnamefont {G.~S.}\ \bibnamefont
  {Beavers}}, \bibinfo {author} {\bibfnamefont {K.}~\bibnamefont {Wittenberg}},
  \ and\ \bibinfo {author} {\bibfnamefont {E.~M.}\ \bibnamefont {Sparrow}},\
  }\bibfield  {title} {\enquote {\bibinfo {title} {Fluid flow through a class
  of highly-deformable porous media. {P}art {II}: {E}xperiments with water},}\
  }\href@noop {} {\bibfield  {journal} {\bibinfo  {journal} {Journal of Fluids
  Engineering}\ }\textbf {\bibinfo {volume} {103}},\ \bibinfo {pages}
  {440--444} (\bibinfo {year} {1981})}\BibitemShut {NoStop}%
\bibitem [{\citenamefont {Parker}\ \emph {et~al.}(1987)\citenamefont {Parker},
  \citenamefont {Mehta},\ and\ \citenamefont {Caro}}]{parker1987steady}%
  \BibitemOpen
  \bibfield  {author} {\bibinfo {author} {\bibfnamefont {K.~H.}\ \bibnamefont
  {Parker}}, \bibinfo {author} {\bibfnamefont {R.~V.}\ \bibnamefont {Mehta}}, \
  and\ \bibinfo {author} {\bibfnamefont {C.~G.}\ \bibnamefont {Caro}},\
  }\bibfield  {title} {\enquote {\bibinfo {title} {Steady flow in porous,
  elastically deformable materials},}\ }\href@noop {} {\bibfield  {journal}
  {\bibinfo  {journal} {Journal of Applied Mechanics}\ }\textbf {\bibinfo
  {volume} {54}},\ \bibinfo {pages} {794--800} (\bibinfo {year}
  {1987})}\BibitemShut {NoStop}%
\bibitem [{\citenamefont {Hewitt}\ \emph {et~al.}(2016)\citenamefont {Hewitt},
  \citenamefont {Nijjer}, \citenamefont {Worster},\ and\ \citenamefont
  {Neufeld}}]{hewitt2016flow}%
  \BibitemOpen
  \bibfield  {author} {\bibinfo {author} {\bibfnamefont {Duncan~R}\
  \bibnamefont {Hewitt}}, \bibinfo {author} {\bibfnamefont {Japinder~S}\
  \bibnamefont {Nijjer}}, \bibinfo {author} {\bibfnamefont {M~Grae}\
  \bibnamefont {Worster}}, \ and\ \bibinfo {author} {\bibfnamefont {Jerome~A}\
  \bibnamefont {Neufeld}},\ }\bibfield  {title} {\enquote {\bibinfo {title}
  {Flow-induced compaction of a deformable porous medium},}\ }\href@noop {}
  {\bibfield  {journal} {\bibinfo  {journal} {Physical Review E}\ }\textbf
  {\bibinfo {volume} {93}},\ \bibinfo {pages} {023116} (\bibinfo {year}
  {2016})}\BibitemShut {NoStop}%
\bibitem [{\citenamefont {MacMinn}\ \emph {et~al.}(2016)\citenamefont
  {MacMinn}, \citenamefont {Dufresne},\ and\ \citenamefont
  {Wettlaufer}}]{macminn2016large}%
  \BibitemOpen
  \bibfield  {author} {\bibinfo {author} {\bibfnamefont {Christopher~W}\
  \bibnamefont {MacMinn}}, \bibinfo {author} {\bibfnamefont {Eric~R}\
  \bibnamefont {Dufresne}}, \ and\ \bibinfo {author} {\bibfnamefont {John~S}\
  \bibnamefont {Wettlaufer}},\ }\bibfield  {title} {\enquote {\bibinfo {title}
  {Large deformations of a soft porous material},}\ }\href@noop {} {\bibfield
  {journal} {\bibinfo  {journal} {Physical Review Applied}\ }\textbf {\bibinfo
  {volume} {5}},\ \bibinfo {pages} {044020} (\bibinfo {year}
  {2016})}\BibitemShut {NoStop}%
\bibitem [{\citenamefont {Preziosi}\ \emph {et~al.}(1996)\citenamefont
  {Preziosi}, \citenamefont {Joseph},\ and\ \citenamefont
  {Beavers}}]{preziosi1996infiltration}%
  \BibitemOpen
  \bibfield  {author} {\bibinfo {author} {\bibfnamefont {L.}~\bibnamefont
  {Preziosi}}, \bibinfo {author} {\bibfnamefont {D.~D.}\ \bibnamefont
  {Joseph}}, \ and\ \bibinfo {author} {\bibfnamefont {G.~S.}\ \bibnamefont
  {Beavers}},\ }\bibfield  {title} {\enquote {\bibinfo {title} {Infiltration of
  initially dry, deformable porous media},}\ }\href@noop {} {\bibfield
  {journal} {\bibinfo  {journal} {International Journal of Multiphase Flow}\
  }\textbf {\bibinfo {volume} {22}},\ \bibinfo {pages} {1205--1222} (\bibinfo
  {year} {1996})}\BibitemShut {NoStop}%
\bibitem [{\citenamefont {MacMinn}\ \emph {et~al.}(2015)\citenamefont
  {MacMinn}, \citenamefont {Dufresne},\ and\ \citenamefont
  {Wettlaufer}}]{macminn2015fluid}%
  \BibitemOpen
  \bibfield  {author} {\bibinfo {author} {\bibfnamefont {C.~W.}\ \bibnamefont
  {MacMinn}}, \bibinfo {author} {\bibfnamefont {E.~R.}\ \bibnamefont
  {Dufresne}}, \ and\ \bibinfo {author} {\bibfnamefont {J.~S.}\ \bibnamefont
  {Wettlaufer}},\ }\bibfield  {title} {\enquote {\bibinfo {title} {Fluid-driven
  deformation of a soft granular material},}\ }\href@noop {} {\bibfield
  {journal} {\bibinfo  {journal} {Physical Review X}\ }\textbf {\bibinfo
  {volume} {5}},\ \bibinfo {pages} {011020} (\bibinfo {year}
  {2015})}\BibitemShut {NoStop}%
\bibitem [{\citenamefont {Hencky}(1931)}]{hencky1931law}%
  \BibitemOpen
  \bibfield  {author} {\bibinfo {author} {\bibfnamefont {H.}~\bibnamefont
  {Hencky}},\ }\bibfield  {title} {\enquote {\bibinfo {title} {The law of
  elasticity for isotropic and quasi-isotropic substances by finite
  deformations},}\ }\href {\doibase 10.1122/1.2116361} {\bibfield  {journal}
  {\bibinfo  {journal} {Journal of Rheology}\ }\textbf {\bibinfo {volume}
  {2}},\ \bibinfo {pages} {169--176} (\bibinfo {year} {1931})}\BibitemShut
  {NoStop}%
\bibitem [{\citenamefont {Anand}(1979)}]{anand1979h}%
  \BibitemOpen
  \bibfield  {author} {\bibinfo {author} {\bibfnamefont {L.}~\bibnamefont
  {Anand}},\ }\bibfield  {title} {\enquote {\bibinfo {title} {On {H. H}encky's
  approximate strain-energy function for moderate deformations},}\ }\href@noop
  {} {\bibfield  {journal} {\bibinfo  {journal} {Journal of Applied Mechanics}\
  }\textbf {\bibinfo {volume} {46}},\ \bibinfo {pages} {78--82} (\bibinfo
  {year} {1979})}\BibitemShut {NoStop}%
\bibitem [{\citenamefont {Bazant}(1998)}]{bazant1998easy}%
  \BibitemOpen
  \bibfield  {author} {\bibinfo {author} {\bibfnamefont {Z.~P.}\ \bibnamefont
  {Bazant}},\ }\bibfield  {title} {\enquote {\bibinfo {title} {Easy-to-compute
  tensors with symmetric inverse approximating {H}encky finite strain and its
  rate},}\ }\href@noop {} {\bibfield  {journal} {\bibinfo  {journal} {Journal
  of Engineering Materials and Technology}\ }\textbf {\bibinfo {volume}
  {120}},\ \bibinfo {pages} {131--136} (\bibinfo {year} {1998})}\BibitemShut
  {NoStop}%
\bibitem [{\citenamefont {Xiao}\ and\ \citenamefont
  {Chen}(2002)}]{xiao2002hencky}%
  \BibitemOpen
  \bibfield  {author} {\bibinfo {author} {\bibfnamefont {H.}~\bibnamefont
  {Xiao}}\ and\ \bibinfo {author} {\bibfnamefont {L.~S.}\ \bibnamefont
  {Chen}},\ }\bibfield  {title} {\enquote {\bibinfo {title} {Hencky's
  elasticity model and linear stress-strain relations in isotropic finite
  hyperelasticity},}\ }\href@noop {} {\bibfield  {journal} {\bibinfo  {journal}
  {Acta Mechanica}\ }\textbf {\bibinfo {volume} {157}},\ \bibinfo {pages}
  {51--60} (\bibinfo {year} {2002})}\BibitemShut {NoStop}%
\bibitem [{\citenamefont {Trefethen}(2000)}]{trefethen2000spectral}%
  \BibitemOpen
  \bibfield  {author} {\bibinfo {author} {\bibfnamefont {L.~N.}\ \bibnamefont
  {Trefethen}},\ }\href@noop {} {\emph {\bibinfo {title} {Spectral Methods in
  {MATLAB}}}}\ (\bibinfo  {publisher} {SIAM},\ \bibinfo {year}
  {2000})\BibitemShut {NoStop}%
\bibitem [{\citenamefont {Pich{\'e}}\ and\ \citenamefont
  {Kanniainen}(2007)}]{piche2007solving}%
  \BibitemOpen
  \bibfield  {author} {\bibinfo {author} {\bibfnamefont {Robert}\ \bibnamefont
  {Pich{\'e}}}\ and\ \bibinfo {author} {\bibfnamefont {Juho}\ \bibnamefont
  {Kanniainen}},\ }\bibfield  {title} {\enquote {\bibinfo {title} {Solving
  financial differential equations using differentiation matrices.}}\ }in\
  \href@noop {} {\emph {\bibinfo {booktitle} {World Congress on Engineering}}}\
  (\bibinfo {year} {2007})\ pp.\ \bibinfo {pages} {1016--1022}\BibitemShut
  {NoStop}%
\bibitem [{\citenamefont {Bj{\o}rnar{\aa}}\ and\ \citenamefont
  {Mathias}(2013)}]{bjornaraa2013pseudospectral}%
  \BibitemOpen
  \bibfield  {author} {\bibinfo {author} {\bibfnamefont {Tore~I}\ \bibnamefont
  {Bj{\o}rnar{\aa}}}\ and\ \bibinfo {author} {\bibfnamefont {Simon~A}\
  \bibnamefont {Mathias}},\ }\bibfield  {title} {\enquote {\bibinfo {title} {A
  pseudospectral approach to the {McWhorter and Sunada Equation} for two-phase
  flow in porous media with capillary pressure},}\ }\href@noop {} {\bibfield
  {journal} {\bibinfo  {journal} {Computational Geosciences}\ }\textbf
  {\bibinfo {volume} {17}},\ \bibinfo {pages} {889--897} (\bibinfo {year}
  {2013})}\BibitemShut {NoStop}%
\bibitem [{\citenamefont {Weideman}\ and\ \citenamefont
  {Reddy}(2000)}]{weideman2000matlab}%
  \BibitemOpen
  \bibfield  {author} {\bibinfo {author} {\bibfnamefont {J.~A.~C.}\
  \bibnamefont {Weideman}}\ and\ \bibinfo {author} {\bibfnamefont {S.~C.}\
  \bibnamefont {Reddy}},\ }\bibfield  {title} {\enquote {\bibinfo {title} {A
  {MATLAB} differentiation matrix suite},}\ }\href@noop {} {\bibfield
  {journal} {\bibinfo  {journal} {ACM Transactions on Mathematical Software
  (TOMS)}\ }\textbf {\bibinfo {volume} {26}},\ \bibinfo {pages} {465--519}
  (\bibinfo {year} {2000})},\ \bibinfo {note} {the codes are available at
  {http://dip.sun.ac.za/$\sim$weideman/research/differ.html}}\BibitemShut
  {NoStop}%
\end{thebibliography}
%

\clearpage
\appendix

\section{Hencky elasticity \textit{vs}. linear elasticity for a uniaxial deformation}\label{supp:Hencky}

\begin{figure}[b]
    \centering
    \includegraphics[width=8.6cm]{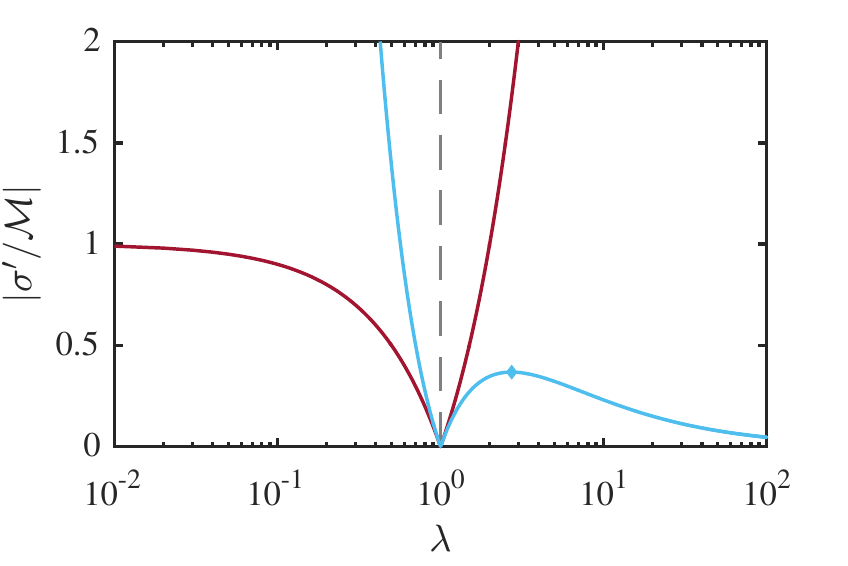}
    \caption{The absolute value of the dimensionless effective stress \textit{vs}. the stretch for uniaxial deformation according to linear elasticity (red) and Hencky elasticity (blue). Hencky elasticity provides a stiffer response than linear elasticity in compression ($\lambda<1$) and a softer response in tension ($\lambda>1$). In tension, the stress predicted by Hencky elasticity reaches a maximum value of $\sigma^\prime/\mathcal{M}=1/e$ for a stretch of $\lambda=e$ (blue diamond) before decreasing asymptotically to zero. The two models agree in the limit of small strain, $|\lambda-1|\ll{}1$ \label{fig:hencky} }
\end{figure}

For a simple uniaxial deformation, Hencky elasticity reduces to
\begin{equation}
    \frac{\sigma^\prime}{\mathcal{M}}=\frac{\ln{\lambda}}{\lambda},
\end{equation}
where $\sigma^\prime$ is the normal effective stress, $\mathcal{M}$ is the oedometric modulus, and $\lambda=1+\Delta{L}/L$ is the stretch, with $\Delta{L}$ the change in overall length and $L$ the original length. Linear elasticity instead predicts
\begin{equation}
    \frac{\sigma^\prime}{\mathcal{M}}=\lambda-1.
\end{equation}
We compare these behaviours in Figure~\ref{fig:hencky}.

\section{Injection via fixed pressure drop}\label{supp:fixed_dp}

To enforce a constant pressure drop $\Delta{p}$, we must derive an associated expression for the evolving flow rate $q(t)$. To do so, we rearrange and integrate the expression for $v_s$ from Eq.~(2.12b) to obtain
\begin{equation}
    q(t) =\frac{\Delta{p}+\displaystyle\int_a^b\,\frac{v_s}{k(\phi_f)}\,\mathrm{d}r}{\displaystyle\int_a^b\,\frac{1}{rk(\phi_f)}\,\mathrm{d}r}
\end{equation}
where
\begin{equation}
    v_s=\frac{\mathrm{D}u_s}{\mathrm{D}t} =\lambda_r\frac{\partial{u_s}}{\partial{t}}.
\end{equation}
For linearised kinematics, we replace the latter expression with
\begin{equation}
    v_s\approx{}\frac{\partial{u_s}}{\partial{t}}.
\end{equation}
The above expressions also apply at steady state, for which $v_s\equiv{}0$.

\section{Linear elasticity with constant permeability (L-$k_0$ and Q-$k_0$)}\label{supp:LQ-k0solutions}

Assuming linear elasticity, we solve Eq.~(3.2) for constant permeability ($k[\phi_f(u_s)] \equiv 1$) to arrive at a general expression for the displacement,
\begin{subequations}\label{genund}
\begin{equation}
    u_{s} =-\frac{qr\ln r}{2}+\frac{(2B_1+q)r}{2(1+\Gamma)}+\frac{B_2}{(1-\Gamma)r},
\end{equation}
where $B_1$ and $B_2$ are determined by the boundary conditions. This result is solely mechanical and constitutive, and is therefore valid for both the L-$k_0$ and Q-$k_0$ models. The general expressions for the effective stresses are then
\begin{equation}\label{LQ-k0_stresses}
    \sigma_r^\prime =-\frac{(1+\Gamma)}{2}q\ln r +B_1-\frac{B_2}{r^2} \quad\mathrm{and}\quad
\sigma_\theta^\prime =-\frac{(1+\Gamma)}{2}q\ln r +B_1+\frac{B_2}{r^2}+\frac{q}{2}(1-\Gamma).
\end{equation}
\end{subequations}
From these expressions, we arrive at four distinct solutions by combining the two different treatments of the kinematics (rigorous Q and linearised L) with the two different sets of outer boundary conditions (an applied stress at the outer boundary (Eq.~2.27) and a fixed outer boundary (Eq.~2.28)). The two L-$k_0$ solutions are classical solutions from linear poroelasticity~[35]. An approximate version of the Q-$k_0$ solutions was derived by Barry and Aldis~[3] and Barry and Mercer~[4], who applied boundary conditions at the moving boundary but linearised the relationship between $\phi_f$ and $u_s$.

\subsection{Solution for L-$k_0$ with an applied effective stress at the outer boundary}\label{supp:Lk0-ap-solution}

For an applied effective stress at the outer boundary, we derive expressions for $B_1$ and $B_2$ by applying the appropriate inner and outer boundary conditions (Eqs.~(2.25) and (2.27), respectively). We linearise the kinematics by applying these at $r=a_0$ (rather than at $a$) and at $r=1$ (rather than at $b$), respectively. We obtain
\begin{equation}
    B_1 = \frac{2\sigma_r^\star-(1+\Gamma )q\ln(a_0)}{2(1-a_0^2)}+\frac{1+\Gamma}{2}q\ln(a_0) \quad \text{and}\quad B_2 = \frac{a^2[2\sigma_r^\star-(1+\Gamma)q\ln(a_0)]}{2(1-a_0^2)}.
\end{equation}

\subsection{Solution for L-$k_0$ with a fixed outer boundary}\label{supp:Lk0-zd-solution}

Similarly, for a fixed outer boundary, we apply the appropriate inner and outer boundary conditions (Eqs.~(2.25) and (2.28), respectively) at $r=a_0$ and at $r=1$, respectively, to obtain
\begin{subequations}
    \begin{equation}
        B_1=-(1-\Gamma)\left\{\frac{q[1+(1+\Gamma) \ln\left(a_0\right)]}{2[a_0^2(1+\Gamma)+(1-\Gamma)]}\right\}+\frac{(1+\Gamma)}{2}q\ln a_0
    \end{equation}
and
    \begin{equation}
        B_2= -(1-\Gamma)\left\{\frac{qa_0^2[1+(1+\Gamma) \ln\left(a_0\right)]}{2[a_0^2(1+\Gamma)+(1-\Gamma)]}\right\}.
    \end{equation}
\end{subequations}
All other quantities can be derived from the expressions for $u_s$. Thus, we have complete explicit solutions following classical linear poroelasticity for the two different sets of outer boundary conditions. Note that, for linearised kinematics, $\phi_f$ should be calculated from $u_s$ according to Eq.~(2.16).

\subsection{Solution for Q-$k_0$ with an applied effective stress at the outer boundary}\label{supp:Qk0-ap-solution}

For an applied effective stress at the outer boundary, we now apply Eqs.~(2.25) and (2.28) at $r=a$ and $r=b$, respectively, to the general elastic solution (Eq.~\ref{genund}). This leads to
\begin{equation}
    B_1 = \frac{b^2[2\sigma_r^\star+(1+\Gamma )q\ln(b/a)]}{2(b^2-a^2)}+\frac{1+\Gamma}{2}q\ln(a), \quad
B_2 = \frac{a^2b^2[2\sigma_r^\star+(1+\Gamma )q\ln(b/a)]}{2(b^2-a^2)}.
\end{equation}
This solution is not explicit because the inner radius $a$ and outer radius $b$ are now determined by the two kinematic conditions (see Eqs.~(2.25) and (2.27)), leading to two coupled, implicit expressions for $a$ and $b$. We solve these expressions numerically using a root-finding technique.

\subsection{Solution for Q-$k_0$ with a fixed outer boundary}\label{supp:Qk0-zd-solution}

For a fixed outer boundary, we now obtain
\begin{subequations} \label{B}
    \begin{equation}
        B_1=-(1-\Gamma)\left\{\frac{q[1+(1+\Gamma) \ln\left(a\right)]}{2[a^2(1+\Gamma)+(1-\Gamma)]}\right\} +\frac{(1+\Gamma)}{2}q\ln a
    \end{equation}
and
    \begin{equation}
        B_2= -(1-\Gamma)\left\{\frac{qa^2[1+(1+\Gamma) \ln\left(a\right)]}{2[a^2(1+\Gamma)+(1-\Gamma)]}\right\}.
    \end{equation}
\end{subequations}
The problem is closed by applying the kinematic condition at the inner boundary (see Eq.~(2.25)), leading to an implicit expression for $a$. We again solve this numerically using a root-finding technique. As above, all other quantities can then be derived from the expressions for $u_s$. Note that, for rigorous kinematics, $\phi_f$ should be calculated from $u_s$ according to Eq.~(2.27).

\section{Numerical solution via Chebyshev spectral collocation}\label{supp:Chebyshev}

When the ODE presented in \S\ref{ststsolns} cannot be solved analytically, it must instead be integrated using standard numerical methods for BVPs, such as direct finite differences or a shooting method. For a shooting method, one must guess the locations of the free boundaries, solve the ODE as an initial value problem (IVP) subject to two of the constraints, and then iterate on the guesses until the remaining constraints are satisfied. For direct finite differences, two approaches are possible. One may follow the same approach as for a shooting method, but solving the BVP directly using finite differences and root finding (\textit{e.g.}, Newton's method) rather than solving it as an IVP. Alternatively, one may solve the BVP and all constraints simultaneously using finite differences and root finding.

\begin{figure}[tb]
    \centering
    \includegraphics[width=5in]{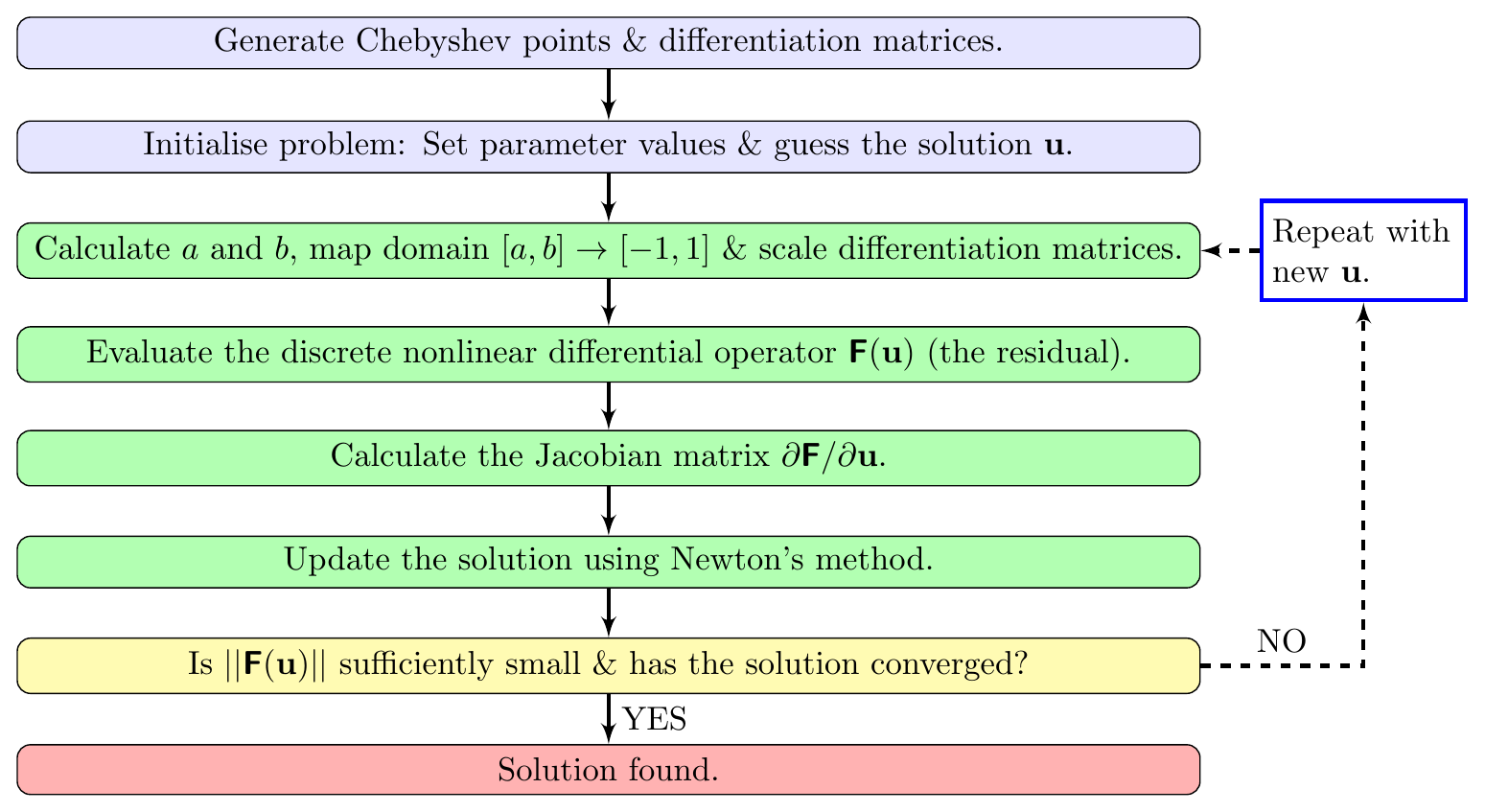}
    \caption{Procedure for direct solution via Chebyshev spectral collocation method. \label{fig:flowchart} }
\end{figure}

Although straightforward to implement, these approaches are unreliable in the present context because the iteration process can easily lead to a nonphysical state that prohibits further iteration. To mitigate these difficulties, we instead use a direct method based on Chebyshev spectral collocation (\textit{i.e.}, a Chebyshev pseudospectral method) [\textit{e.g.}, 8, 21]. That is, we solve the BVP and all constraints simultaneously as described above, but replacing the sparse finite-difference differentiation matrix with a dense Chebyshev-pseudospectral differentiation matrix. This approach still requires Newton iteration, but is more robust than finite differences because the density of the pseudospectral differentiation matrix directly couples the solution at each discrete point to the solution at every discrete point. This approach also allows for the straightforward incorporation of additional unknowns and constraints, such as solving the problem for an imposed pressure drop $\Delta{p}$ rather than for an imposed flow rate $q$. We illustrate the overall structure of the method in Figure~\ref{fig:flowchart}. Note that, for purposes of Newton iteration, we calculate the Jacobian analytically for the L and Q models and numerically for the N models.

Spectral collocation methods involve discretising the solution domain into a set of $N$ points (collocation points), defining a global function that interpolates the solution at these collocation points (the interpolant), and then approximating the derivatives of the solution as the derivatives of the interpolant. In a Chebyshev spectral collocation method, the collocation points are the $N$ Chebyshev points $x_k\in[1,-1]$, which can be defined as~[21]
\begin{equation}\label{eq:Chebpoints}
    x_k = \cos\left(\frac{(k-1)\pi}{N-1}\right), \quad k = 1,\ldots{},N.
\end{equation}
The basis functions from which the interpolant is composed are then a set of $N$ polynomials of degree $N-1$ satisfying the criterion that each is nonzero at exactly one distinct collocation point. Note that other definitions of the Chebyshev points are also commonly used~[\textit{e.g.}, 31]. For the definition given in Eq.~(\ref{eq:Chebpoints}), Weideman~and~Reddy~[36] provide a suite of \verb+MATLAB+ functions that generate the Chebyshev points and differentiation matrices, and that perform interpolation.

\section{Impact of geometry}\label{supp:geo}

\begin{figure*}
    \centering
    \includegraphics[width=5in]{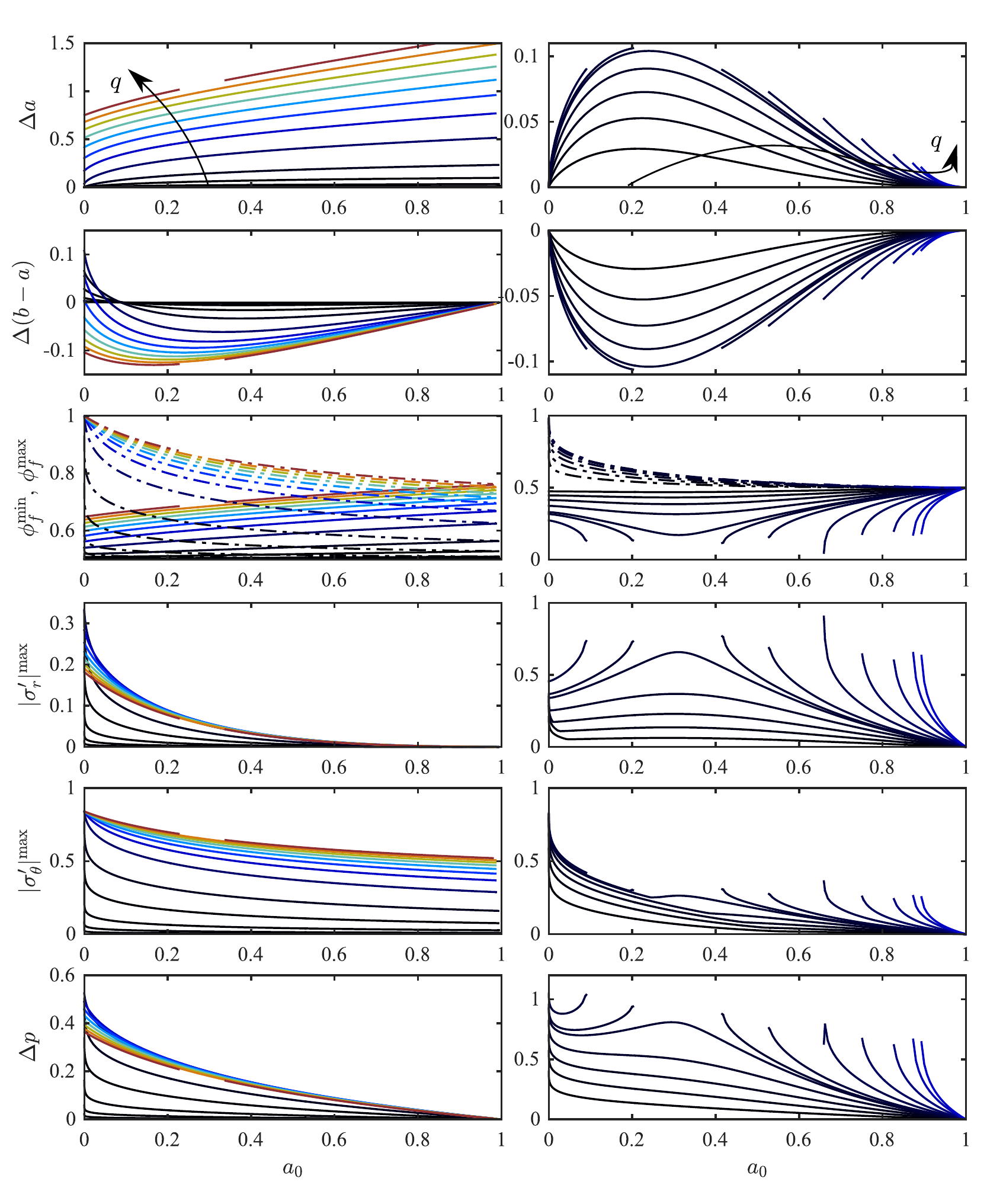}
    \caption{The results of Figure~4 plotted as contours of fixed $q$ against $a_0$, with $q\in{}[0.001,8]$ (left, black to red) and $q\in{}[0.001,2]$ (right, black to blue). \label{fig:geo-a0_q}}
\end{figure*}

\begin{figure*}
    \centering
    \includegraphics[width=5in]{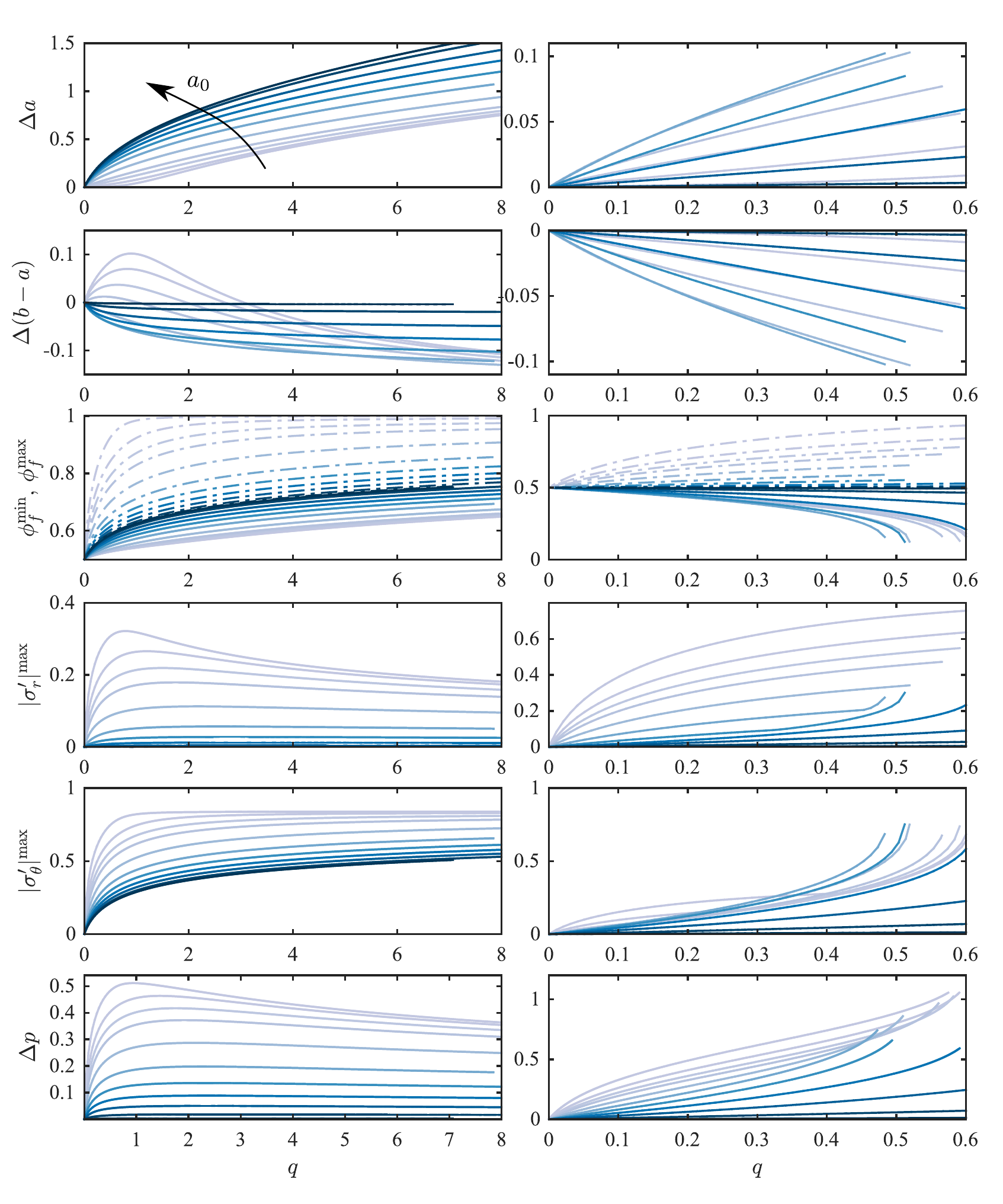}
    \caption{The results of Figure~4 plotted as contours of fixed $a_0$ against $q$, with $a_0\in[0.001,0.98]$ (light to dark). \label{fig:geo-q_a0}}
\end{figure*}

\begin{figure*}
    \centering
    \includegraphics[width=5in]{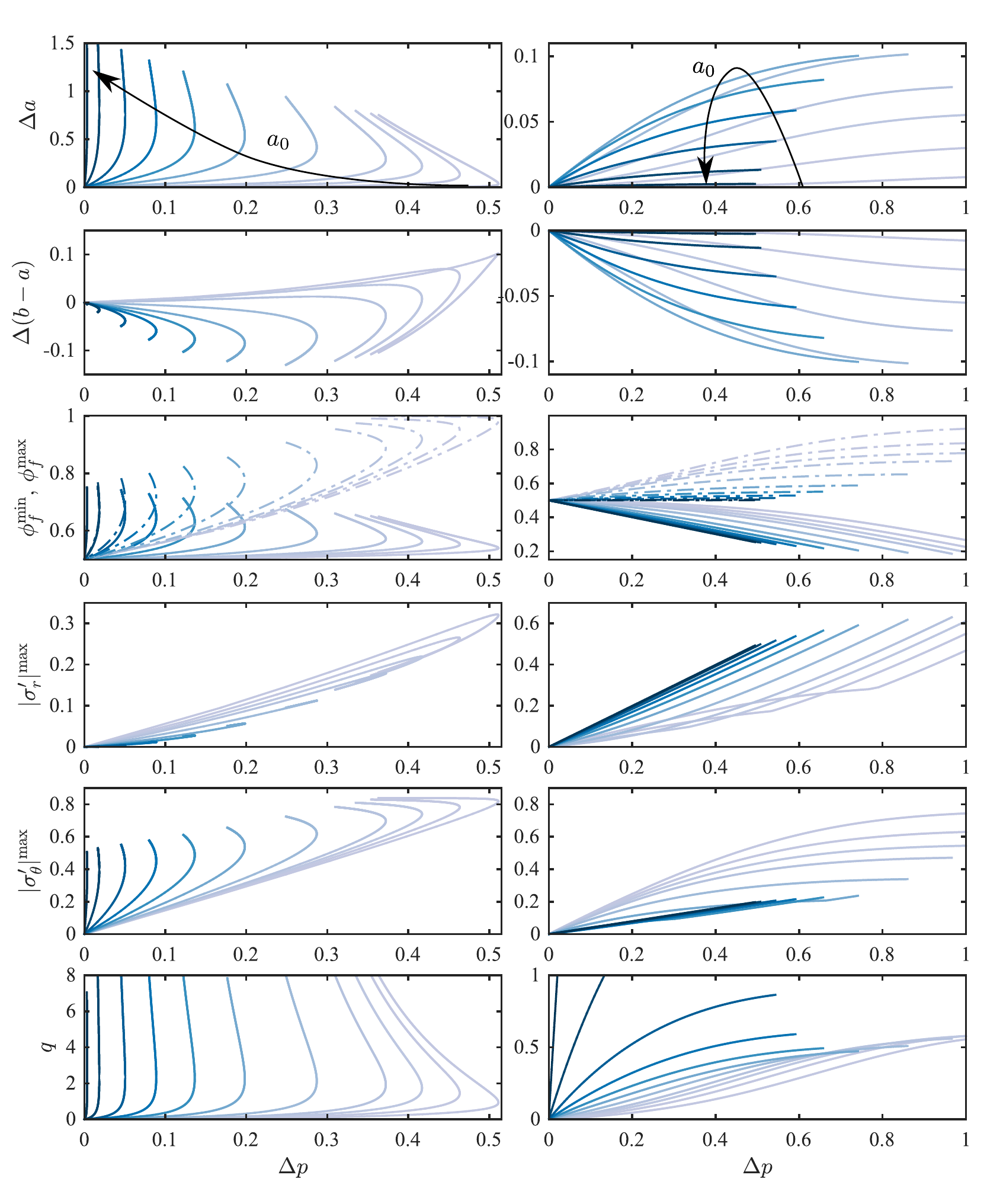}
    \caption{The results of Figure~4 plotted as contours of fixed $a_0$ against $\Delta{p}$, with $a_0\in[0.001,0.98]$ (light to dark). \label{fig:geo-dp_a0}}
\end{figure*}

\begin{figure*}
    \centering
    \includegraphics[width=5in]{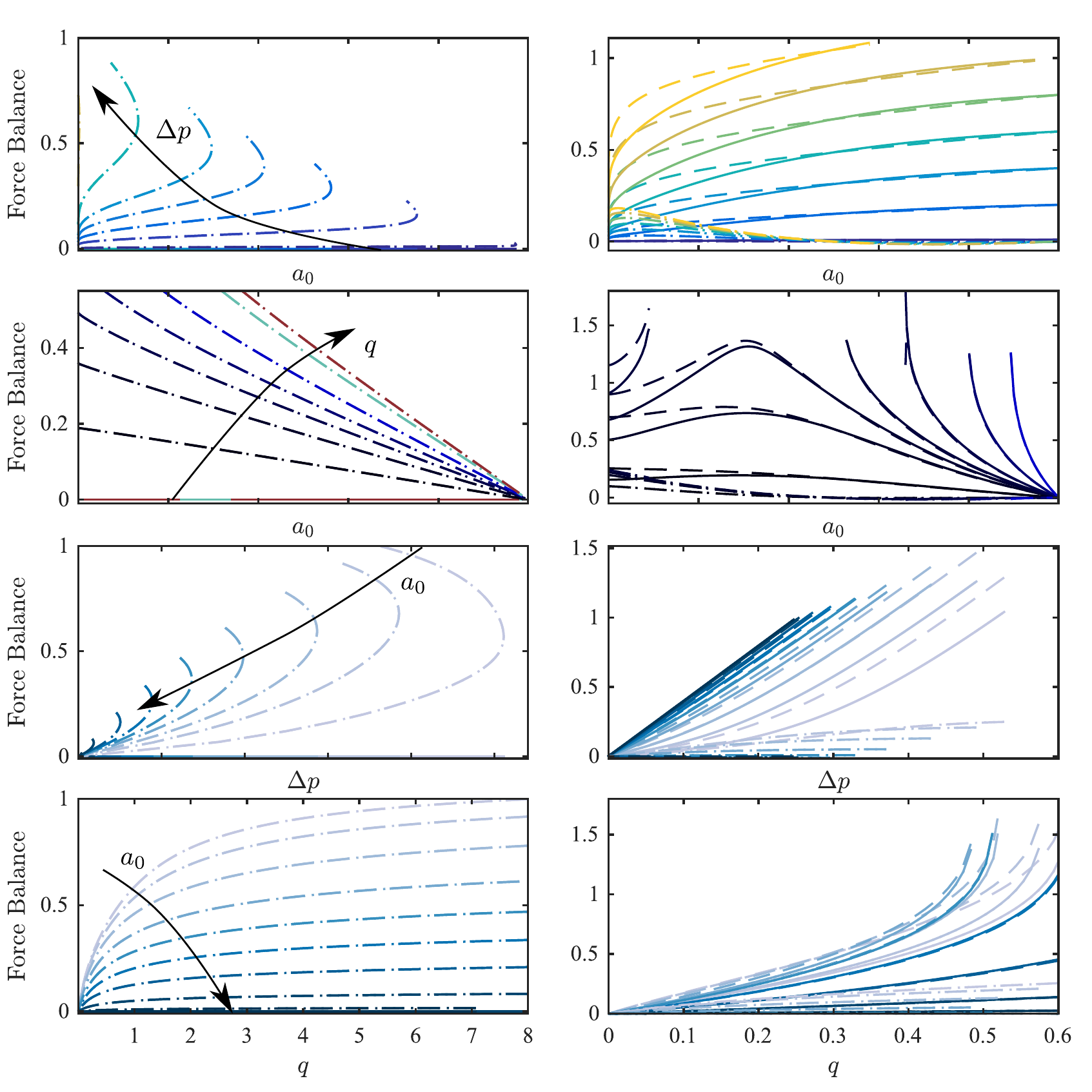}
    \caption{The results of Figure~5 plotted in various combinations. The colours in each row are the same as in the corresponding `version' of Fig.~4. \label{fig:ForceBalanceCombo}}
\end{figure*}

In Figures~4 and 5 of the main text, we plot the evolution of various key quantities as contours of fixed $\Delta{p}$ against $a_0$. It is useful for interpretation to present the same results in several different ways. Here, we show the results of Figure~4 as contours of fixed $q$ against $a_0$ (Fig.~\ref{fig:geo-a0_q}), contours of fixed $a_0$ against $q$ (Fig.~\ref{fig:geo-q_a0}), and contours of fixed $a_0$ against $\Delta{p}$ (Fig.~\ref{fig:geo-dp_a0}). We also do the same for Figure~5 (Fig.~\ref{fig:ForceBalanceCombo}).


\clearpage

\section{Rheological effects}\label{supp:rheo}

\begin{figure*}[b]
    \centering
    \includegraphics[width=5in]{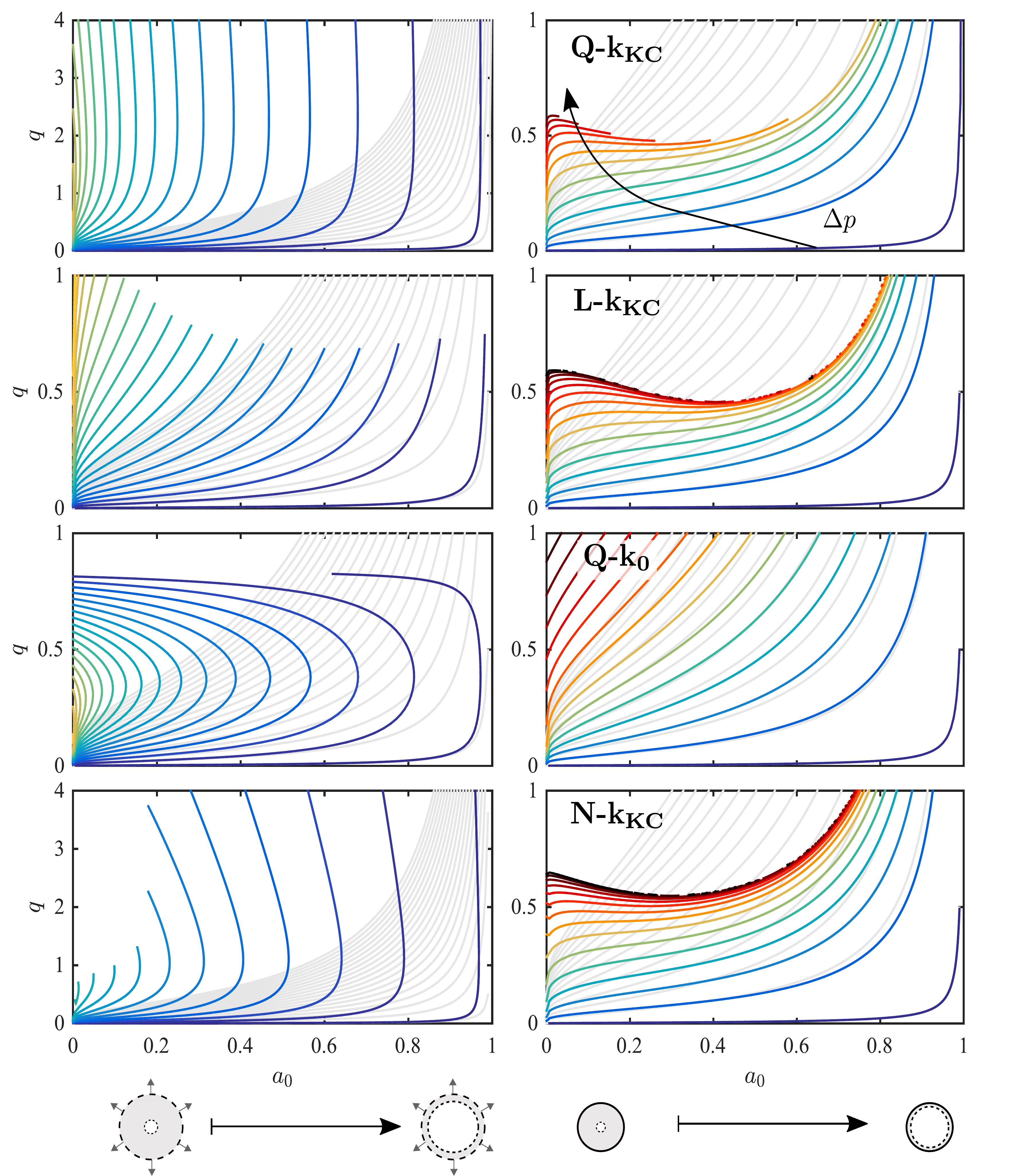}
    \caption{To illustrate the impact of rheology, we plot $q$ against $a_0$ for contours of fixed $\Delta{p}$, as in the last row of Figure~4, for five different models: Q-$k_\mathrm{KC}$ (first row), L-$k_\mathrm{KC}$ (second row), Q-$k_0$ (third row), N-$k_\mathrm{KC}$ (last row), and L-$k_0$ (all rows, grey lines). The colour scale is the same as in Figure~4. \label{modelcomp}}
\end{figure*}

We focused in the main text on results from the Q-$k_\mathrm{KC}$ model because it provides a good compromise between rigour, robustness, and computational efficiency. We examine this choice in more detail in Figure~\ref{modelcomp} by plotting $q$ against $a_0$ for contours of fixed $\Delta{p}$, as in the last row of Figure~4, for five different models: Q-$k_\mathrm{KC}$ (first row), L-$k_\mathrm{KC}$ (second row), Q-$k_0$ (third row), N-$k_\mathrm{KC}$ (last row), and L-$k_0$ (all rows, grey lines). Note that $q$ appears to be much more sensitive to the permeability law than other aspects of the deformation (\textit{cf.} Figures~2 and 3), making it a useful quantity for this comparison.

For unconstrained cylinders (left column), the Q-$k_\mathrm{KC}$ and N-$k_\mathrm{KC}$ models predict qualitatively similar behaviour, with the contours in the latter bending to the left somewhat more strongly. The latter model is also much more computationally expensive. The Q-$k_0$ model exhibits similar behaviour, but with much more extreme bending of the contours (note the different vertical scale). Our results for the L-$k_\mathrm{KC}$ model are inconclusive because this model is much less robust than either of the Q 
models; our method fails to find a solution for even moderate values of $q$. This is likely because the L-$k_\mathrm{KC}$ model is asymptotically inconsistent and does not correctly capture the kinematic relationship between porosity and displacement. The Q-$k_\mathrm{KC}$ is much more rigorous in these regards, and is only slightly more computationally expensive in our pseudospectral collocation framework.

For constrained cylinders (right column), all three of the $k_\mathrm{KC}$ models exhibit very similar behaviour despite the different elasticity laws (L and Q \textit{vs.} N) and the different treatments of the kinematics (L \textit{vs.} Q and N). This presentation does not constitutive a careful quantitative comparison, but it suggests that deformation-dependent permeability plays a key role in the mechanics of the problem, particularly for high pressures (right), whereas large-deformation kinematics are less important. This is somewhat unsurprising since constrained cylinders generally deform much less than unconstrained cylinders.

These results suggest that rigorous large-deformation kinematics (including the relationship between porosity and displacement) are important for model robustness and are central to the double-valued behaviour of unconstrained cylinders. Deformation-dependent permeability appears to moderate (but not eliminate) the double-valued behaviour of unconstrained cylinders, and to be central to the behaviour of constrained cylinders.

As noted in the main text, we have chosen Kozeny-Carman permeability and Hencky elasticity as relatively generic constitutive laws that capture the qualitatively important features of deformation-dependent permeability and large-deformation elasticity, respectively. Given the strong role of deformation-dependent permeability in our results, a comparison with results for other permeability laws would be an interesting topic for future work.

\clearpage

\section{Solution for Q-$k_0$ in the thin-walled limit}\label{supp:thinwall}

We now derive an approximate solution to the Q-$k_0$ model in the limit of vanishing wall thickness, starting from Eq.~(3.2) with $k[\phi_f(u_s)]\equiv1$. We do this for the case of an applied effective stress at the outer boundary since the case of no displacement at the outer boundary is ultimately limited to small displacements and thus is well-captured by linear poroelasticity.

We begin by defining a new radial coordinate $\epsilon\equiv{}r-a$ such that $\epsilon\in[0,\delta]$ where $\delta\equiv{}b-a\ll{}1$ is the wall thickness. We then rewrite Eq.~(3.2) in terms of $\epsilon$ and seek a solution under the assumption that $\epsilon\ll{}1$. From these assumptions, and writing $u_s(r) = U(\epsilon)$, we obtain at leading order
\begin{equation}\label{qpain}
    \frac{\mathrm{d}^2U}{\mathrm{d}\epsilon^2} +\frac{1}{a}\frac{\mathrm{d}U}{\mathrm{d}\epsilon}-\frac{U}{a^2}= -\frac{q}{a}.
\end{equation}
Note that these assumptions require for asymptotic consistency that $q/a=O(1)$.

Equation~\eqref{qpain} is a linear, second-order ODE with solution
\begin{equation}
U = A_1\exp\left[-\left(\frac{\sqrt{5} +1}{2a}\right)\epsilon\right]+A_2\exp\left[\left(\frac{\sqrt{5} -1}{2a}\right)\epsilon\right] +aq.
\end{equation}
We now apply the relevant boundary conditions (Eqs.~(2.25) and (2.27) with $\sigma_r^\star\equiv{}0$), which results in four equations for four unknowns: The two integration constant, $A_1$ and $A_2$, and the inner and outer radii, $a$ and $a+\delta$, respectively. We use the two conditions at the inner boundary to derive expressions for $A_1$ and $A_2$ in terms of $a$,
\begin{equation}
A_1 = \frac{(a-a_0)(\Gamma-1)+q}{\sqrt{5}} \quad\mathrm{and}\quad A_2 = \frac{(\sqrt{5}+1)[a(1-q)-a_0]-\Gamma(a-a_0)}{\sqrt{5}}.
\end{equation}
The two conditions at the outer boundary then give
\begin{subequations}
\begin{equation}
A_1\exp\left[-\left(\frac{\sqrt{5}+1}{2a}\right)\delta\right]+ A_2\exp\left[\left(\frac{\sqrt{5}-1}{2a}\right)\delta\right]+aq = a+\delta-1
\end{equation}
and
\begin{equation}
\left(\frac{\sqrt{5}-1}{2a}\right)A_2\exp\left[\left(\frac{\sqrt{5}-1}{2a}\right)\delta\right] - \left(\frac{\sqrt{5}+1}{2a}\right)A_1\exp\left[-\left(\frac{\sqrt{5}+1}{2a}\right)\delta\right] +\frac{\Gamma}{a}(a+\delta-1) =\sigma_r^\star.
\end{equation}
\end{subequations}
This is now a root-finding problem for the values of $a$ and $\delta$, which we solve numerically using the standard \verb+MATLAB+ function \verb+fsolve+. The pressure field is given by
\begin{equation}
\frac{\mathrm{d}p}{\mathrm{d}r} = -\frac{q}{r} \quad\mapsto\quad \frac{\mathrm{d}P}{\mathrm{d}\epsilon} = -\frac{q}{a} \quad\implies\quad P(\epsilon) = \frac{q}{a}(\delta-\epsilon).
\end{equation}
This then leads to $\Delta{p}=(q/a)\delta$ and, since $q/a=O(1)$, we have that $\Delta{p}=O(\delta)$. This implies that a small pressure drop will drive a large flow rate when the walls are sufficiently thin.


\end{document}